\documentclass[12pt]{article}
\usepackage{a4wide}
\usepackage{epsfig}
\newcommand{\eps}{\varepsilon}

\newcommand{\Disc}{\mbox{\sl Disc\,}}

\newcommand{\Li}{{\rm Li}}
\newcommand{\dDk}{\frac{d^Dk}{(2\pi)^D}}

\newcommand{\dDp}{\frac{d^Dp}{(2\pi)^D}}

\newcommand{\slk}{/\kern-6pt k}
\newcommand{\sll}{/\kern-4pt l}
\newcommand{\slp}{p\kern-5pt/}
\newcommand{\slP}{P\kern-7pt/}
\newcommand{\slq}{q\kern-5.5pt/}
\newcommand{\slv}{v\kern-5pt\raise1pt\hbox{$\scriptstyle/$}\kern1pt}

\newcommand{\pfrac}[2]{\left(\frac{#1}{#2}\right)}
\newcommand{\dalembertian}{\hbox{$\,\vbox{\hrule\hbox{\vrule%
  \vbox{\kern7pt}\kern7pt\vrule}\hrule}\,$}}

\begin{document}

\thispagestyle{empty}
\begin{flushright}
MZ-TH/03-10\\
hep-ph/0307290\\
July 2003\\
\end{flushright}
\vspace{0.5cm}
\begin{center}
{\Large\bf Lectures on configuration space methods for sunrise-type
  diagrams}\\[12pt]
{\large Stefan Groote}\\[7pt]
Institut f\"ur Physik der Johannes-Gutenberg-Universit\"at,\\[3pt]
Staudinger Weg 7, 55099 Mainz, Germany\\[12pt]
{\em Calc2003, Dubna, June 16--18th, 2003}
\end{center}
\begin{abstract}\noindent
In this lecture series I will give a fundamental insight into configuration
space techniques which are of help to calculate a broad class of Feynman
diagrams, the sunrise-type diagrams. Applications are shown along with basic
concepts and techniques.
\end{abstract}

\tableofcontents

\newpage

\section{Introduction}
There has been a renewal of interest in calculating the loop integral of a
simple topology, the so-called generalized sunrise (or sunset, water melon,
banana, basketball) diagrams~\cite{Grozin:2002nb,Mastrolia:2002gt,
Mastrolia:2002tv,Czyz:2002re,Schroder:2002re}. The two-loop sunrise diagrams
with different values of internal masses have been recently studied in some
detail (see e.g.~\cite{Berends:1993ee,
Post:1997dk,Post:1996gg,Rajantie:1996cw,Berends:1997vk,Gasser:1998qt} and
references therein). In this lecture series I outline the configuration space
technique~\cite{Mendels:wc,Groote:1998ic,Groote:1998wy,Groote:1999cx,
Groote:2000kz,Mendels:qe,Bashir:2001ad,Delbourgo:2003zi}. It can be used to
verify the results obtained within other techniques~\cite{Caffo:2002wm,
Caffo:2002ch,Ligterink:1999mu,Fleischer:1999mp} and to investigate some
general questions of Feynman diagram calculation~\cite{Czyz:2002re,
Groote:1999cx,Broadhurst:1991fi,Groote:1999cn,Laporta:2001dd,Passarino:2001wv,
Suzuki:2000wj,Witten:1979kh,Davydychev:2000na}.
The diagram with the sunrise topology form a subset of the general set of
diagram with a given number of loops~\cite{Chetyrkin:1996ia}. They appear in
various specific physics applications:
\begin{itemize}

\item sum rules for baryonic
currents~\cite{Groote:1999zp,Pivovarov:1991nk,Ovchinnikov:1991mu,Groote:2000py,
Grozin:1992td,Furnstahl:1995nd,Jin:1997pb,Kras}

\item gluonic
correlators~\cite{Groote:2001vr,Kataev:1982gr,Pivovarov:1999mr}

\item multiloop calculations~\cite{Kazakov:km,Grozin:2002nb}

\item decays with multi-particle phase-space~\cite{Bardin:1994sc}

\item lattice QCD calculations~\cite{Wetzorke:2002mx}

\item mixing of neutral mesons~\cite{Narison:1994zt}

\item Chiral perturbation theory (ChPT) and effective theories for Goldstone
modes in higher orders of momentum
expansion~\cite{Weinberg:1978kz,Post:1997dk,Gasser:1998qt}

\item exotics~\cite{Larin:1986yt,Sakai:1999qm}

\item effective potentials for symmetry breaking~\cite{effpot,Jackiw:1980kv}

\item finite temperature calculations~\cite{Gross:1980br,Jackiw:1980kv,
hotQCD,Rajantie:1996cw,Kajantie:2003ax,Andersen:2000zn,Nishikawa:2003js}

\item applications in nuclear physics~\cite{nuclph}

\item applications in solid state physics~\cite{Platter:2002yr}

\end{itemize}

\section{Concepts of configuration space techniques}
\begin{figure}[ht]\begin{center}
\epsfig{figure=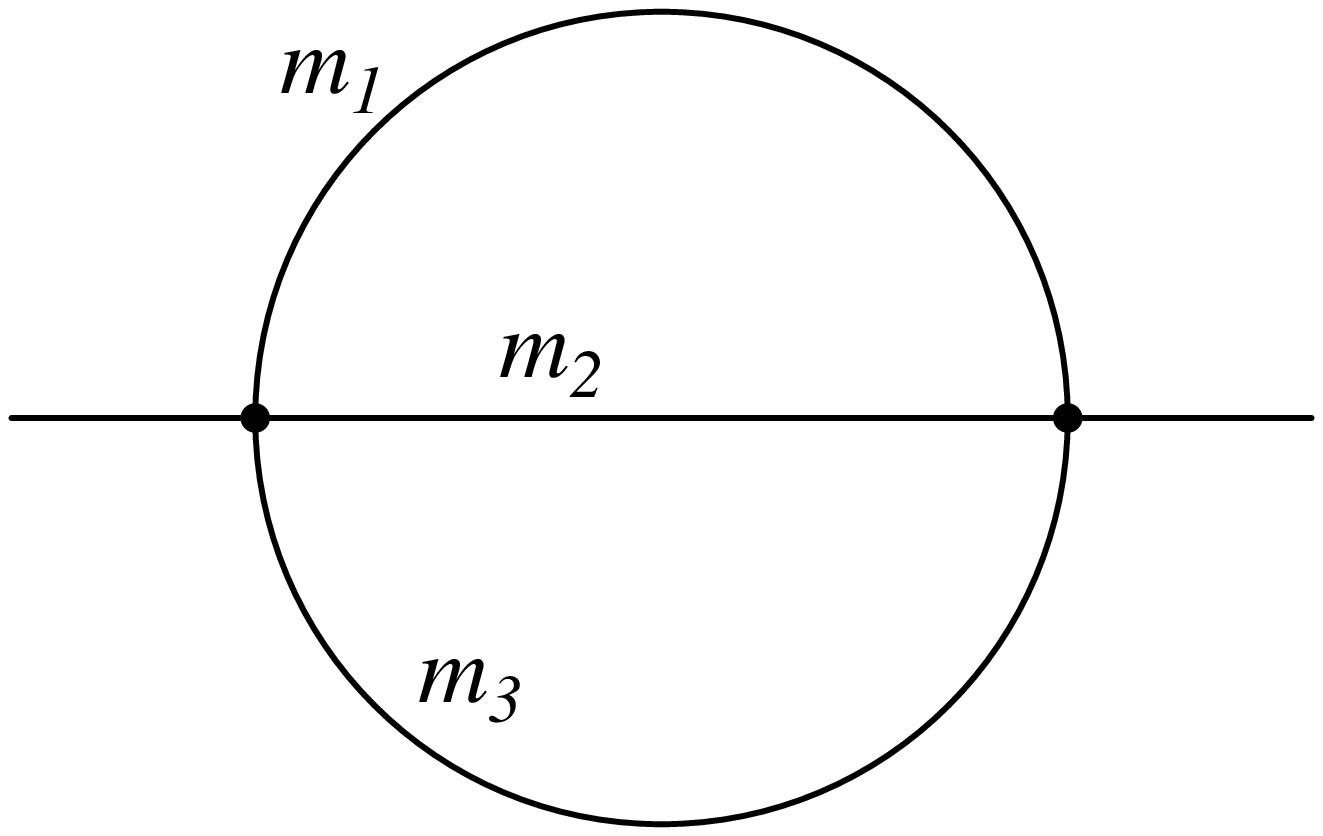, scale=0.3}\qquad
\epsfig{figure=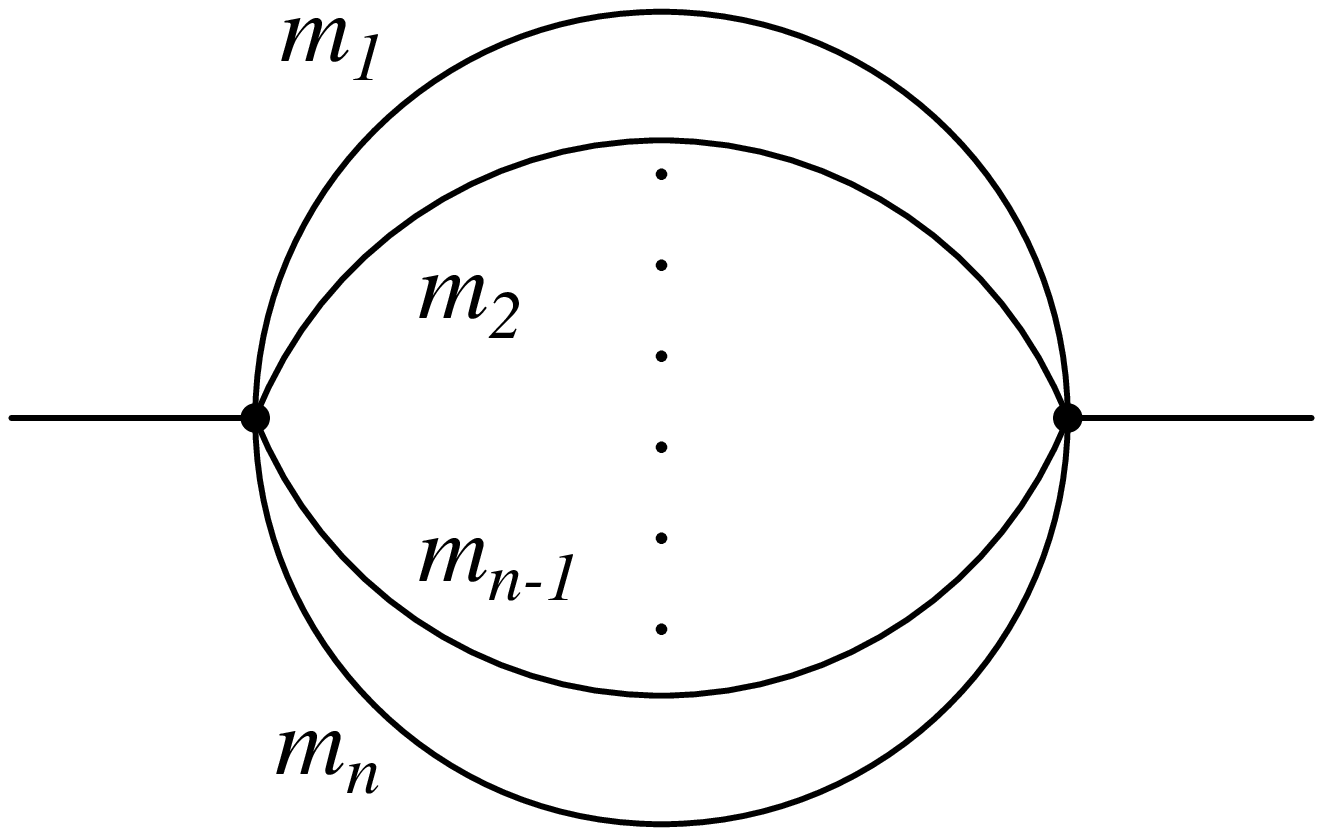, scale=0.3}
\caption{\label{fig01}Genuine sunrise and general topology of the class of
sunrise-type diagrams}\end{center}
\end{figure}

I will start this lecture series with explaining the basic elements that
occur in calculating sunrise-type diagrams. The genuine sunrise diagram with
three internal lines is shown on the left hand side in Fig.~\ref{fig01}. It
is the leading order perturbative correction to the lowest order propagator in
$\phi^4$-theory. The corresponding leading order perturbative correction in
$\phi^3$-theory is a one-loop diagram and can be considered as a degenerate
case of the prior example. A straightforward generalization of this topology
is a correction to the free propagator in $\phi^{n+1}$-theory that contains
$(n-1)$ loops and $n$ internal lines (see the right hand side of
Fig.~\ref{fig01}).

In configuration space, the ``calculation'' of the correlator two-point
function is very easy. It is just given by the product of the propagators,
\begin{equation}
\Pi(x)=\prod_iD(x,m_i).
\end{equation}
The propagators themselves (for the scalar case and massive particles) are
solutions of the Klein--Gordon equation and can be written in $D$-dimensional
Euclidean space-time as
\begin{equation}
D(x,m)=\int\dDp\frac{e^{ip_\mu x^\mu}}{p^2+m^2}
  =\frac{(mx)^\lambda K_\lambda(mx)}{(2\pi)^{\lambda+1}x^{2\lambda}}
  =\frac{(m/x)^\lambda}{(2\pi)^{\lambda+1}}K_\lambda(mx)
\end{equation}
where I used $D=2\lambda+2$ for convenience, $x=\sqrt{x_\mu x^\mu}$, and
$K_\lambda(z)$ is the modified Bessel function of the second kind (see
e.g.~\cite{Watson}). Frankly speaking, the propagator $D(x,m)$ describes the
propagation of a particle of mass $m$ from the space-time point $0$ to $x$ (a
situation which of course is invariant with respect to translation). For $m=0$
we obtain the massless propagator
\begin{equation}
D(x,0)=\int\dDp\frac{e^{ik_\mu x^\mu}}{p^2}
  =\frac{\Gamma(\lambda)}{4\pi^{\lambda+1}x^{2\lambda}}
\end{equation}
simply by calculating the limit, taking into account the series expansion of
the Bessel function $K_\lambda(mx)$ which will be shown later on. Modified
internal lines can also occur, as a result of a reduction procedure of more
complicated diagrams or as diagrams with mass insertion on a line. This
insertions are indicated by one or more dots placed on the line. Again, the
representation of these cases is easily established,
\begin{equation}
D^{(\mu)}(x,m)=\int\dDp\frac{e^{ip_\mu x^\mu}}{(p^2+m^2)^{\mu+1}}
  =\frac{(m/x)^{\lambda-\mu}}{(2\pi)^{\lambda+1}2^\mu\Gamma(\mu+1)}
  K_{\lambda-\mu}(mx)
\end{equation}
(note that $D^{(\mu)}(x,m)$ is related to the $\mu$-th derivative with respect
to the parameter $m^2$). It is obvious that, up to different power in $x$ and
normalization, a modification of the internal line just changes the index of
the Bessel function. The modified propagator has the same functional form
as the usual propagator. This matter of fact unifies and simplifies the
considerations.

The generalization to more complicated integrands with additional tensor
structure $p^{\mu_1}\cdots p^{\mu_k}$ is straightforward as well. Because of
\begin{equation}
\int\dDp\frac{p_\nu e^{ip_\mu x^\mu}}{p^2+m^2}
  =-i\frac\partial{\partial x^\nu}\int\dDp\frac{e^{ip_\mu x^\mu}}{p^2+m^2}
  =-i\frac{x_\nu}{x}\frac\partial{\partial x}
  \pfrac{(mx)^\lambda K_\lambda(mx)}{(2\pi)^{\lambda+1}x^{2\lambda}},
\end{equation}
this merely leads again to different orders of the Bessel function, together
with a tensor structure $x^{\mu_1}\cdots x^{\mu_k}$. However, these tensor
structure does not matter in taking simply the product of the propagators.
We can conclude that in order to obtain the expression for the correlator
function in configuration space there is no integration at all.

\subsection{The correlator function in momentum space}
An integration occurs, if we want to calculate the correlator function in
momentum space, being the Fourier transform of $\Pi(x)$,
\begin{equation}
\tilde\Pi(p)=\int\Pi(x)e^{ip_\mu x^\mu}d^Dx.
\end{equation}
For a general tensor structure, the integral is calculated by expanding the
plane wave function $\exp(ip_\mu x^\mu)$ in a series of Gegenbauer polynomials
$C_j^\lambda(p_\mu x^\mu/px)$. The Gegenbauer polynomials are a generalizations
of associated Legendre polynomials to the $D$-dimensional space-time. They are
determined by the characteristic polynomial
\begin{equation}
(t^2-2tz+1)^{-\lambda}=\sum_{l=0}^\infty t^lC_l^\lambda(z)
\end{equation}
and are given by $C_0^\lambda(z)=1$, $C_1^\lambda(z)=2\lambda z$ and the
recursion formula
\begin{equation}
(l+1)C_{l+1}^\lambda(z)=2(l+\lambda)zC_l^\lambda(z)
  -(l+2\lambda-1)C_{l-1}^\lambda(z)=0.
\end{equation}
The Gegenbauer polynomials satisfy the orthogonality relations
($\hat x_i=x_i/|x_i|$)
\begin{equation}
\int C_m^\lambda(\hat x_1\cdot\hat x_2)C_n^\lambda(\hat x_2\cdot x_3)
  d^D\hat x_2=\frac{2\pi^{\lambda+1}}{\Gamma(\lambda+1)}
  \frac{\lambda\delta_{mn}}{\lambda+n}C_n^\lambda(\hat x_1\cdot\hat x_3)
\end{equation}
on the $D$-dimensional unit sphere with rotationally invariant measure
$d^D\hat x_2$ where
\begin{equation}
\int d^D\hat x_2=\frac{2\pi^{\lambda+1}}{\Gamma(\lambda+1)}.
\end{equation}
Expressed in terms of Gegenbauer polynomials, the plane wave reads
\begin{equation}
\exp(ip_\mu x^\mu)=\Gamma(\lambda)\pfrac{px}2^{-\lambda}
  \sum_{l=0}^\infty i^l(\lambda+l)J_{j+l}(px)C_l^\lambda(p_\mu x^\mu/px)
\end{equation}
where $J_\lambda(z)$ is the Bessel function of the first kind and
$p=\sqrt{p_\mu p^\mu}$. This formula allows one to single out an irreducible
tensorial structure from the angular integration in the Fourier integral.
Integration techniques involving Gegenbauer polynomials for the computation of
massless diagrams are described in Detail in~\cite{Chetyrkin:pr} where many
useful relations can be found (see also~\cite{Terrano:1980af}). In case of
absence of explicit tensorial structure, however, the orthogonality relations
for Gegenbauer polynomials can be used to explicitely integrate over the unit
sphere,
\begin{equation}
\int e^{ip_\mu x^\mu}d^D\hat x=2\pi^{\lambda+1}\pfrac{px}2^\lambda
  J_\lambda(px).
\end{equation}
In the following I will concentrate on this scalar case. The Fourier transform
of the sunrise-type diagram, therefore, is given by the one-dimensional
integral
\begin{equation}
\tilde\Pi(p)=2\pi^{\lambda+1}\int_0^\infty\pfrac{px}2^{-\lambda}
  J_\lambda(px)D(x,m_1)\cdots D(x,m_n)x^{2\lambda+1}dx.
\end{equation}

\subsection{The renormalization}
While the configuration space expression for the sunrise-type diagram contains
no integration, it does not always represent an integrable function for a
generic value of space-time dimension $D$. $\Pi(x)$ can have non-integrable
singularities at small $x$ for a sufficiently large number of propagators when
$D>2$~\cite{Bogoliubov}. The reason is that each propagator by itself can be
considered as a distribution. However, distributions do not form an algebra
and multiplication is not well-defined. In this respect the correlator function
$\Pi(x)$ is not completely defined as a proper distribution. In attempting to
integrate such a function over the whole $x$-space we encounter infinities
which are usual UV divergencies. Therefore, the computation of its Fourier
transform requires regularization (for instance, dimensional regularization)
and subtraction.

UV divergencies given as poles in $\eps$ for the dimensional regularization
are related to short distances $x$. Therefore, one should expand massive
propagators at small $x$ to obtain the counterterms. Still a convergence at
large $x$ should be retained which provides IR regularization (see
e.g.~\cite{Pivovarov:re}). This technical constraint can be met for instance
by keeping one (or two) massive propagators unexpanded. The possibility is
given because the integrals of one or two Bessel functions are
known~\cite{Prudnikov,Gradshteyn}.

However, one can do even simpler by using a damping function like
$e^{-\mu^2x^2}$ to suppress contributions for large values of $x$,
corresponding to IR singularities. This is done by introducing the factor
$1=e^{\mu^2x^2}e^{-\mu^2x^2}$ into the integrand. In this case all massive
propagators along with the factor $e^{\mu^2x^2}$ can be expanded in small
values of $x$. The expansion of the propagators is straightforward and results
from the definitions
\begin{equation}
J_\lambda(z)=\pfrac z2^\lambda\sum_{k=0}^\infty
  \frac{(-z^2/4)^k}{k!\Gamma(\lambda+k+1)}
  =\frac{(z/2)^\lambda}{\Gamma(1+\lambda)}\left(1-\frac{(z/2)^2}{1+\lambda}
  +\ldots\ \right)
\end{equation}
and
\begin{equation}
K_\lambda(z)=\frac\pi2\frac{I_{-\lambda}(z)-I_\lambda(z)}{\sin(\pi\lambda)},
  \qquad\Gamma(\lambda)\Gamma(1-\lambda)=\frac\pi{\sin(\pi\lambda)}
\end{equation}
with
\begin{equation}
I_\lambda(z)=\pfrac z2^\lambda
  \sum_{k=0}^\infty\frac{(z^2/4)^k}{k!\Gamma(\lambda+k+1)}
\end{equation}
Therefore, one has 
\begin{equation}\label{Kexpand}
\left(\frac z2\right)^\lambda K_\lambda(z)
  =\frac{\Gamma(\lambda)}2\left[1+\frac1{1-\lambda}\left(\frac z2\right)^2
  -\frac{\Gamma(1-\lambda)}{\Gamma(1+\lambda)}
  \left(\frac z2\right)^{2\lambda}\right]+O(z^4,z^{2+2\lambda}).
\end{equation}
The final integration can be done formally by using the identity
\begin{equation}
\int_0^\infty x^{r-1}e^{-\mu^2x^2}dx=\frac12\mu^{-r}\Gamma(r/2).
\end{equation}
As an example we consider the genuin sunrise diagram with three different
masses $m_1$, $m_2$, and $m_3$, given in configuration space by
\begin{equation}
\Pi(x)=D(x,m_1)D(x,m_2)D(x,m_3).
\end{equation}
The counterterms for the Fourier transform
\begin{equation}
\tilde\Pi(p)=2\pi^{\lambda+1}\int_0^\infty\pfrac{px}2^{-\lambda}
  J_\lambda(px)D(x,m_1)D(x,m_2)D(x,m_3)x^{2\lambda+1}dx
\end{equation}
are obtained by introducing $e^{\mu^2x^2}e^{-\mu^2x^2}$ and expanding the
occuring Bessel functions as well as $e^{\mu^2x^2}$. The singular parts we
obtain for dimensional regularization with $D=4-2\eps$, i.e.\
$\lambda=1-\eps$, are given by
\begin{equation}
\tilde\Pi_{\rm sing}(p)=\frac{\mu^{-4\eps}}{\pi^{4-2\eps}}
  \Bigg\{-\frac{m_1^2+m_2^2+m_3^2}{512\eps^2}
  +\frac1\eps\left(\sum_{i=1}^3\frac{m_i^2\ln(m_ie^{\gamma\prime}/2\mu)}{128}
  -\frac{p^2}{1024}\right)\Bigg\}
\end{equation}
where $\gamma'=\gamma_E/2-3/4$. Note, though, that the result is given in
Euclidean domain. In this way one easily finds pole parts for any diagram with
any mass arrangement.

\subsection{The momentum subtraction}
Renormalization by momentum subtraction is the oldest renormalization method.
The idea is to subtract the integrand in some specific momentum point. For
massive diagrams the point $p=0$ is IR-safe, and the receipe is realized for
instance by expanding the function
\begin{equation}
\pfrac{px}2^{-\lambda}J_\lambda(px)
\end{equation}
which is the kernel or weight function of the integral transformation) in a
Taylor series around $p=0$ in terms of a polynomial series in $p^2$. The
subtraction of order $N$ is achieved by writing
\begin{equation}
\left[\left(\frac{px}2\right)^{-\lambda}J_\lambda(px)\right]_N
  =\left(\frac{px}2\right)^{-\lambda}J_\lambda(px)
  -\sum_{k=0}^N\frac{(-1)^k}{k!\Gamma(\lambda+k+1)}
  \left(\frac{px}2\right)^{2k}
\end{equation}
and by keeping $N$ terms in the expansion on the right hand side. The number
$N$ of necessary subtractions is determined by the divergence index of the
diagram and can be found according to the standard rules of the
$R$-operation~\cite{Bogoliubov}. For the example we started with, a momentum
subtraction which is symmetric in the masses $m_1$, $m_2$, and $m_3$ leads to
the finite part
\begin{equation}
\tilde\Pi_{\rm fin}(p)=2\pi^{\lambda+1}\int_0^\infty
  \left[\pfrac{px}2^{-\lambda}J_\lambda(px)\right]_ND(x,m_1)D(x,m_2)D(x,m_3)
  x^{2\lambda+1}e^{\mu^2x^2}e^{-\mu^2x^2}dx
\end{equation}
where $N$ is the order of expansion which has to be chosen for integrability.
This is of course the same order which has been used earlier in order to
extract the counterterms. It is obvious, though, that the expansion can cease
at positive powers of $x$. Finally, $\eps=0$ can be taken to calculate the
finite part. As a technical remark note that most numerical integration
routines will run into problems if the upper limit is kept to be infinity.
However, alread for values of the order $x\sim 1$ the integrand is
negligible, so that the integration can terminate at this point. Also the
region about the origin might cause trouble. In this case the integration
interval can be subdivided close to the origin, and the whole integrand for
the subinterval to zero can be expanded in $x$ as a whole. For specific
values of $p^2$, $m_1$, $m_2$, and $m_3$ we could reproduce results given in
the literature (e.g.\ results given in~\cite{Caffo:2002ch,SonDo}).

\subsection{A second example}
The second example I want to show here starts with the same genuine sunrise
diagram but is evaluated for a special setting of the masses. The reason is
that there exists an analytical expression in the
literature~\cite{Berends:1997vk}. In this example the subtraction is done in
several steps. The first step consists of a momentum subtraction for the
weight function with $N=1$. We obtain
\begin{eqnarray}
\tilde\Pi(p)&=&\tilde\Pi_{\rm mom}(p)+\tilde\Pi_{\rm rem}(p)
  \qquad\mbox{where}\\[3pt]
\tilde\Pi_{\rm mom}(p)&=&2\pi^{\lambda+1}\int_0^\infty
  \left[\pfrac{px}2^{-\lambda}J_\lambda(px)\right]_1D(x,m_1)D(x,m_2)D(x,m_3)
  x^{2\lambda+1}dx\ =\nonumber\\&=&2\pi^{\lambda+1}\int_0^\infty
  \left[\pfrac{px}2^{-\lambda}J_\lambda(px)-\frac1{\Gamma(\lambda+1)}
  +\frac{p^2x^2}4\frac1{\Gamma(\lambda+2)}\right]\
  \times\nonumber\\&&\qquad\times\ D(x,m_1)D(x,m_2)D(x,m_3)x^{2\lambda+1}dx,
  \\[7pt]
\tilde\Pi_{\rm rem}(p)&=&A+p^2B\ =\nonumber\\[3pt]
  &=&\frac{2\pi^{\lambda+1}}{\Gamma(\lambda+1)}\int_0^\infty
  D(x,m_1)D(x,m_2)D(x,m_3)x^{2\lambda+1}dx\,+\nonumber\\&&\qquad
  -p^2\frac{2\pi^{\lambda+1}}{4\Gamma(\lambda+2)}\int_0^\infty
  x^2D(x,m_1)D(x,m_2)D(x,m3)x^{2\lambda+1}dx.
\end{eqnarray}
The analytical result in~\cite{Berends:1997vk} is given at the pseudothreshold
$p=m_1+m_2-m_3$. For simplicity we choose $m_1=m_2=m_3/2=m$. In this case we
obtain $p=0$, $\tilde\Pi_{\rm mom}(0)=0$ and the vanishing of the counterterm
$p^2B$. Only
\begin{equation}
A=\frac{2\pi^{\lambda+1}}{\Gamma(\lambda+1)}\int_0^\infty D(x,m)D(x,m)D(x,2m)
  x^{2\lambda+1}dx
\end{equation}
has to be determined. In order to separate singular and finite part as
$(2\pi)^DA=S+F$ (where the total normalization of~\cite{Berends:1997vk} has
been adopted) we use momentum subtraction for the last Bessel function
$K_\lambda(2mx)$ and obtain for the singular part
\begin{eqnarray}
S&=&\frac{(2\pi)^D m^{2\lambda}}{\Gamma(\lambda+1)}
  \int_0^\infty x^{1-2\lambda}K_\lambda(mx)K_\lambda(mx)
  \frac{\Gamma(\lambda)}2\left[1+\frac{(mx)^2}{1-\lambda}
  -\frac{\Gamma(1-\lambda)}{\Gamma(1+\lambda)}(mx)^{2\lambda}\right]dx
  \ =\nonumber\\
&=&\pi^{4-2\eps}\frac{m^{2-4\eps}\Gamma^2(1+\eps)}
  {(1-\eps)(1-2 \eps)}\left[-\frac3{\eps^2}
  +\frac{8\ln 2}{\eps}+8(2-2\ln 2-\ln^2 2)\right]+O(\eps)
\end{eqnarray}
where we could make use of
\begin{equation}\label{KKint}
\int_0^\infty x^{2\alpha-1}K_\mu(mx)K_\mu(m x)dx
  =\frac{2^{2\alpha-3}}{m^{2\alpha}\Gamma(2\alpha)}
  \Gamma(\alpha+\mu)\Gamma(\alpha)\Gamma(\alpha)\Gamma(\alpha-\mu).
\end{equation}
In comparing the part $S$ with the analytical result in~\cite{Berends:1997vk}
for $p=0$,
\begin{equation}
\tilde\Pi_{\rm ref}(0)=\pi^{4-2\eps}\frac{m^{2-4\eps}
  \Gamma^2(1+\eps)}{(1-\eps)(1-2\eps)}\left[-\frac3{\eps^2}
  +\frac{8\ln 2}\eps-8\ln^2 2\right]+O(\eps),
\end{equation}
one obtains the same singular parts while the difference of the finite parts
is expected to be found in the part $F$. Indeed, one obtains
\begin{eqnarray}
F&=&\frac{(2\pi)^Dm^{2\lambda}}{\Gamma(\lambda+1)}
  \int_0^\infty x^{2(1-\lambda)-1}K_\lambda(mx)K_\lambda(m x)\
  \times\nonumber\\&&\times\ \left\{(mx)^\lambda K_\lambda(2mx)
  -\frac{\Gamma(\lambda)}2\left[1+\frac{(mx)^2}{1-\lambda}
  -\frac{\Gamma(1-\lambda)}{\Gamma(1+\lambda)}(mx)^{2\lambda}\right]
  \right\}dx
\end{eqnarray}
which is finite and can therefore be calculated for $D=4$,
\begin{equation}
F=16\pi^4m^2\int_0^\infty\frac{dx}xK_1(x)K_1(x)
  \left\{xK_1(2x)-\frac12\left[1+x^2(-1+2\gamma_E+2\ln x)\right]\right\}.
\end{equation}
It is shown numerically that $F=16\pi^4m^2(\ln 2-1)$, and because this
quantity is finite, the normalization can be restored, we obtain
\begin{equation}
F=\pi^{4-2\eps}\frac{m^{2-4\eps}\Gamma^2(1+\eps)}
  {(1-\eps)(1-2\eps)}16(\ln 2-1)+O(\eps).
\end{equation}

\subsection{A third example}
An example for the three-loop case is the sunrise-type diagram with two
massive and two massless lines at vanishing external momentum. The analytical
expression for the diagram in configuration space representation is given by
\begin{equation}
\tilde\Pi(0)=\int\left(D(x,m)\right)^2\left(D(x,0)\right)^2d^Dx=\int
  \left(\frac{(mx)^\lambda K_\lambda(mx)}{(2\pi)^{\lambda+1}x^{2\lambda}}
  \right)^2\left(\frac{\Gamma(\lambda)}{4\pi^{\lambda+1}x^{2\lambda}}
  \right)^2 d^Dx.
\end{equation}
While the angular integration in $D$-dimensional space-time is trivial,
\begin{equation}
\int e^{ip_\mu x^\mu}d^D\hat x=2\pi^{\lambda+1}\pfrac{px}2^{-\lambda}
  J_\lambda(px)\rightarrow \frac{2\pi^{\lambda+1}}{\Gamma(\lambda+1)}\quad
  \mbox{for $p\rightarrow 0$},
\end{equation}
the problem of residual radial integration is solved by Eq.~(\ref{KKint}).
The result reads
\begin{equation}
\tilde\Pi(0)=\left(\frac{m^2}4\right)^{3\lambda-1}\frac1{2^8\pi^{3\lambda+3}}
\frac{\Gamma(\lambda)^2\Gamma(1-\lambda)\Gamma(1-2\lambda)^2
  \Gamma(1-3\lambda)}{\Gamma(\lambda+1)\Gamma(2-4\lambda)}.
\end{equation}
This result corresponds to the quantity $M_1$ in~\cite{Broadhurst:1991fi}
which is the simplest basis element for the computation of massive three-loop
diagrams in a general three-loop topology considered
in~\cite{Broadhurst:1991fi}. We obtain agreement with~\cite{Broadhurst:1991fi}.

\subsection{The spectral density}
From the physical point of view the interesting part of the analysis of
sunrise-type diagrams is the construction of the spectral decomposition of the
diagrams. For the two-point correlation function we determine the discontinuity
across the physical cut in the complex plane of the squared momentum,
$p^2=-m^2\pm i0$. Taking for instance the $n=2$ sunrise-type diagram with
two different masses $m_1$, $m_2$ in arbitrary dimensions, in (Euclidean)
momentum space given by
\begin{eqnarray}
\tilde\Pi(p)&=&\int\dDk\frac1{(k^2+m_1^2)((k+p)^2+m_2^2)}\ =\ \ldots\nonumber\\
  &=&\frac{\Gamma(1-\lambda)}{(4\pi)^{\lambda+1}}\int_0^1
  \left(x(1-x)p^2-(1-x)m_1^2+xm_2^2\right)^{\lambda-1}dx,
\end{eqnarray}
the physical cut has influence on the integral for
$x(1-x)p^2+(1-x)m_1^2+xm_2^2<0$. With $p^2=-se^{\mp i\pi}$ we can calculate
the discontinuity to obtain
\begin{equation}
{\rm Disc}\left(x(1-x)p^2+(1-x)m_1^2+xm_2^2\right)^{\lambda-1}
  =\frac{2\pi i\left(x(1-x)s-(1-x)m_1^2-xm_2^2
  \right)^{\lambda-1}}{\Gamma(\lambda)\Gamma(1-\lambda)}
\end{equation}
for $x_1\le x\le x_2$ where $x_1$ and $x_2$ are the two zeros of the
integrand which can be decomposed into
$s^{\lambda-1}(x_2-x)^{\lambda-1}(x-x_1)^{\lambda-1}$. The spectral density
is given by the discontinuity divided by $2\pi i$. Therefore, we finally obtain
\begin{equation}\label{rhom1m2}
\rho(s)=\frac{\Omega_{2\lambda+1}}{4\sqrt s(2\pi)^{2\lambda+1}}
  \pfrac{(s-m_1^2-m_2^2)^2-4m_1^2m_2^2}{4s}^{\lambda-1/2}
\end{equation}
where $\Omega_d=2\pi^{d/2}/\Gamma(d/2)$ is the volume of the unit sphere in
$d$-dimensional space-time. Eq.~(\ref{rhom1m2}) is of need in the following
which is the reason why I have given some details about the calculation. In
the third lecture we will see that configuration space technique allows for
some other efficient tool for determining the spectral density, namely by
inverting the so-called $K$-transform. In the next lecture I will continue
with today's examples and give recurrence relations which allow to reduce the
result to a finite set of master integrals, expressible in terms of
transcendental numbers.

\section{Recurrence relations and transcendental numbers}
I continue the previous lecture by introducing just three other examples for
calculating an sunrise-type diagram with the help of configuration space
techniques in order to compare them with results in the literature. A wide
field of diagrams to compare with is a subclass of three-loop bubbles
$B_N$~\cite{Broadhurst:1991fi}, namely
\begin{eqnarray}
\lefteqn{B_N(0,0,n_3,n_4,n_5,n_6)\ =\ \int\frac{d^Dk\ d^Dl\ d^Dp}{m^{3D}
  (\pi^{D/2}\Gamma(3-D/2))^3}\ \times}\nonumber\\&&\times\
  \frac{m^{2n_3}}{((p+k)^2+m^2)^{n_3}}\frac{m^{2n_4}}{((p+l)^2+m^2)^{n_4}}
  \frac{m^{2n_5}}{((p+k+l)^2+m^2)^{n_5}}\frac{m^{2n_6}}{(p^2+m^2)^{n_6}}
\end{eqnarray}
with two propagators absent ($n_1=n_2=0$). Actually we choose the remaining
indices $n_3$, $n_4$, $n_5$, and $n_6$ in a way that the results become finite,
but we not only calculate the finite part but also the part porportional to
$\eps$ in order to compare with~\cite{Broadhurst:1991fi}. Written in
configuration space, the particular subset of bubble diagrams $B_N$ is given by
\begin{eqnarray}
\lefteqn{B_N(0,0,n_3,n_4,n_5,n_6)\ =\ 
  \frac{2(64\pi^4)^{2-\eps}}{(\Gamma(1+\eps))^3\Gamma(2-\eps)}
  m^{2(n_3+n_4+n_5+n_6)-12+6\eps}\ \times}\nonumber\\&&\times\
  \int_0^\infty D^{(n_3-1)}(x,m)D^{(n_4-1)}(x,m)D^{(n_5-1)}(x,m)
  D^{(n_6-1)}(x,m)x^{2\lambda+1}dx.
\end{eqnarray}
It is obvious that we end up with integrals of products of four Bessel
functions with non-integer indices and a non-integer power of $x$. The plan
for today's lecture is as follows: After introducing the results for the three
examples in terms of these integrals, I will show that at least the expansion
up to first order in $\eps$ is possible analytically. I will then introduce a
reduction procedure in order to reduce the integrals to a basic set of two
master integrals,
\begin{equation}
L_4(r):=\int_0^\infty\left(K_0(\xi)\right)^4\xi^rd\xi\qquad\mbox{and}\qquad
L_4^l(r):=\int_0^\infty\left(K_0(x)\right)^4\xi^r\ln(e^{\gamma_E}\xi/2)\,dx
\end{equation}
and will add considerations on the general class of basic integrals $L_n(r)$
and $L_n^l(r)$.

\begin{figure}[ht]\begin{center}
\epsfig{figure=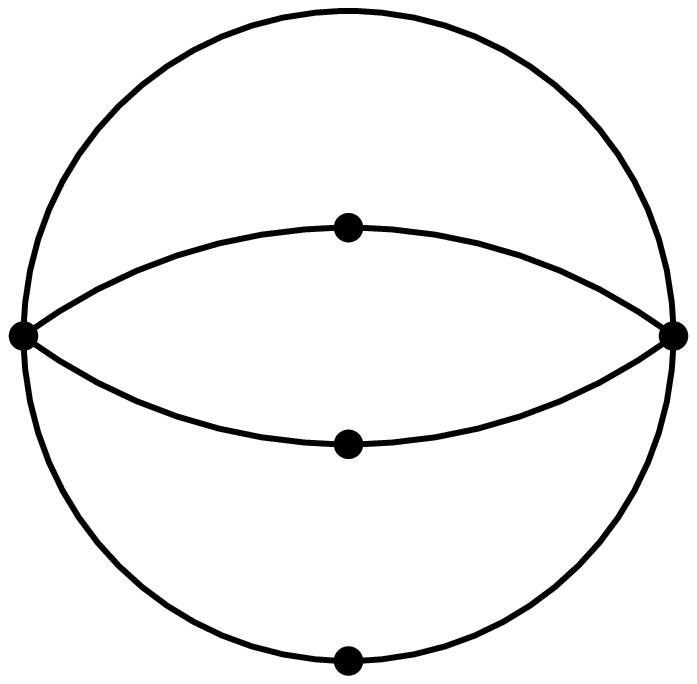, scale=0.5}\qquad
\epsfig{figure=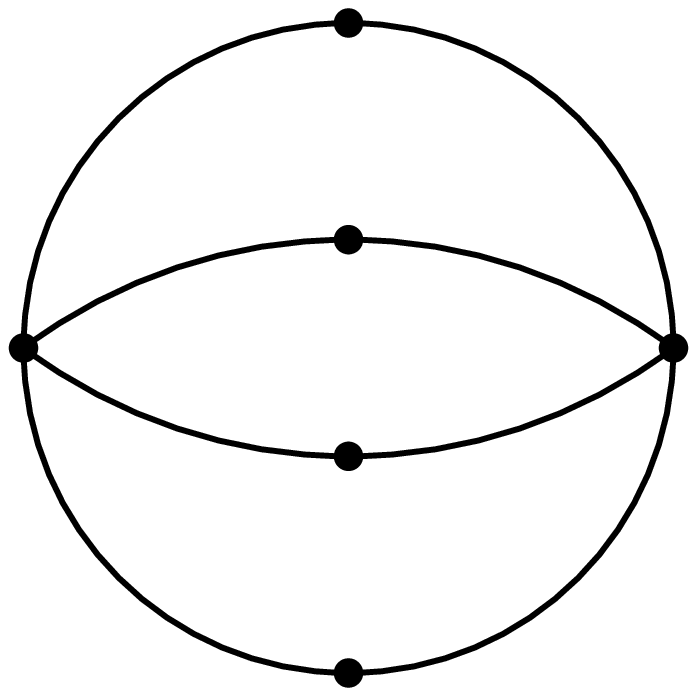, scale=0.5}\qquad
\epsfig{figure=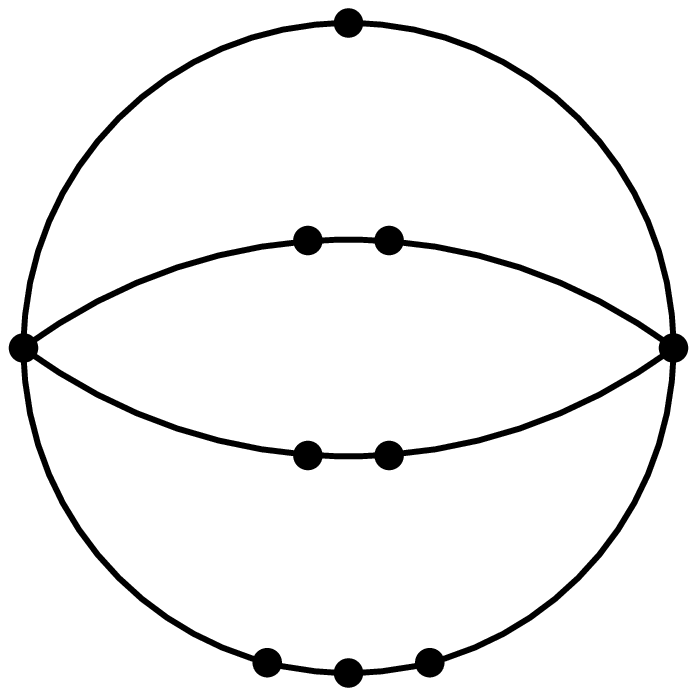, scale=0.5}
\caption{\label{fig02}The diagrams for $B_N(0,0,2,2,2,1)$, $B_N(0,0,2,2,2,2)$,
  and $B_N(0,0,2,3,3,4)$.}\end{center}

\end{figure}

\subsection{The example $B_N(0,0,2,2,2,2)$}
We start with the example which is represented by the central diagram in
Fig.~\ref{fig02}. Each of the lines is modified (indicated by the dots on the
lines) which means that instead of the propagators $D(x,m)$ we have to use
\begin{equation}
D'(x,m)=\int\dDp\frac{e^{ip_\mu x^\mu}}{(p^2+m^2)^2}
  =\frac{(x/m)^{1-\lambda}}{2(2\pi)^{\lambda+1}}K_{\lambda-1}(mx)
  =\frac{(x/m)^\eps}{2(2\pi)^{2-\eps}}K_{-\eps}(mx)
\end{equation}
and obtain ($\xi=mx$)
\begin{equation}
B_N(0,0,2,2,2,2)=\frac{2^{1-2\eps}}{(\Gamma(1+\eps))^3\Gamma(2-\eps)}
  \int_0^\infty\left(K_{-\eps}(\xi)\right)^4\xi^{3+2\eps}d\xi.
\end{equation}
We now can use the general formula~\cite{Gradshteyn}
\begin{equation}\label{indder}
\left[\frac{\partial K_\nu(z)}{\partial\nu}\right]_{\nu=\pm n}
  =\pm\frac12n!\sum_{k=0}^{n-1}\pfrac z2^{k-n}\frac{K_k(z)}{k!(n-k)},\qquad
  n\in\{0,1,\ldots\}
\end{equation}
to expand the Bessel function in a series with respect to its index. In
case of $K_{-\eps}(z)$, however, the first derivative vanishes and we obtain
$K_{-\eps}(z)=K_0(z)+O(\eps^2)$. Therefore, in expanding
\begin{eqnarray}
\frac{2^{1-2\eps}\left(K_{-\eps}(\xi)\right)^4\xi^{3+2\eps}}{(\Gamma(1+\eps)
  )^3\Gamma(2-\eps)}
  &=&2\left(K_0(\xi)\right)^4\xi^3\left(1+(1+2\gamma_E-2\ln 2+2\ln\xi)\eps
  +O(\eps^2)\right)\ =\nonumber\\
  &=&2\left(K_0(\xi)\right)^4\xi^3\left(1+\left(1+2\ln(e^{\gamma_E}\xi/2)
  \right)\eps+O(\eps^2)\right)
\end{eqnarray}
we obtain
\begin{eqnarray}
B_N(0,0,2,2,2,2)&=&2(1+\eps)\int_0^\infty\left(K_0(\xi)\right)^4\xi^3d\xi
  +4\eps\int_0^\infty\left(K_0(\xi)\right)^4\xi^3\ln(e^{\gamma_E}\xi/2)d\xi
  \ =\nonumber\\&=&2(1+\eps)L_4(3)+4\eps L_4^l(3).
\end{eqnarray}
In comparing with the analytical result in~\cite{Broadhurst:1991fi} we
checked numerically~\cite{Groote:1999cx}
\begin{eqnarray}
I_4(3)&=&-\frac3{16}+\frac7{32}\zeta(3),\nonumber\\
I_4^l(3)&=&\frac3{32}+\frac34\Li_4\pfrac12-\frac{17\pi^4}{1920}
  -\frac{\pi^2}{32}\left(\ln 2\right)^2+\frac1{32}\left(\ln 2\right)^4
  +\frac{49}{128}\zeta(3).
\end{eqnarray}

\subsection{The example $B_N(0,0,2,2,2,1)$}
In the diagram on the left hand side of Fig.~\ref{fig02} one of the lines is
not modified. Therefore, we have to deal with one regular propagator factor
\begin{equation}
D^{(0)}(x,m)=D(x,m)=\frac{(x/m)^{-\lambda}}{(2\pi)^{\lambda+1}}K_\lambda(mx)
  =\frac{(x/m)^{\eps-1}}{(2\pi)^{2-\eps}}K_{1-\eps}(mx).
\end{equation}
In this case we obtain
\begin{equation}
B_N(0,0,2,2,2,1)=\frac{2^{2-2\eps}}{(\Gamma(1+\eps))^3\Gamma(2-\eps)}
  \int_0^\infty\left(K_{-\eps}(\xi)\right)^3K_{1-\eps}(\xi)\xi^{2+2\eps}d\xi.
\end{equation}
Using Eq.~(\ref{indder}), we obtain
\begin{equation}
K_{1-\eps}(\xi)=K_1(\xi)-\frac\eps\xi K_0(\xi)+O(\eps^2)
\end{equation}
and
\begin{equation}
\frac{2^{2-2\eps}\left(K_{-\eps}(\xi)\right)^3K_{1-\eps}(\xi)
  \xi^{2+2\eps}}{(\Gamma(1+\eps))^3\Gamma(2-\eps)}
  =4\xi^2\Big(1+\Big(1-\frac1\xi+\ln(e^{\gamma_E}\xi/2)\Big)\eps
  +O(\eps^2)\Big).
\end{equation}
The result reads
\begin{eqnarray}
B(0,0,2,2,2,1)&=&4(1+\eps)\int_0^\infty\left(K_0(\xi)\right)^3K_1(\xi)\xi^2d\xi
  -4\eps\int_0^\infty\left(K_0(\xi)\right)^4\xi\,d\xi\,+\nonumber\\&&\qquad
  +8\eps\int_0^\infty\left(K_0(\xi)\right)^3K_1(\xi)\xi^2
  \ln(e^{\gamma_E}\xi/2)d\xi.
\end{eqnarray}
This result is not yet written in terms of $L_4(r)$ and $L_4^l(r)$ and will be
treated after having introduced the reduction procedure.

\subsection{The example $B_N(0,0,2,3,3,4)$}
In order to demonstrate the power of the configuration space technique also in
this calculation, we finally choose the example found on the right hand side
of Fig.~\ref{fig02}, where the modified propagators
\begin{equation}
D^{(2)}(x,m)=\frac{(x/m)^{1+\eps}}{8(2\pi)^{2-\eps}}K_{-1-\eps}(mx),\qquad
D^{(3)}(x,m)=\frac{(x/m)^{2+\eps}}{48(2\pi)^{2-\eps}}K_{-2-\eps}(mx)
\end{equation}
are used, together with the expansions
\begin{equation}
K_{-1-\eps}(\xi)=K_1(\xi)+\frac\eps\xi K_0(\xi)+O(\eps^2),\qquad
K_{-2-\eps}(\xi)=K_2(x)+\frac{2\eps}\xi K_1(x)+\frac{2\eps}{\xi^2}K_0(x)
  +O(\eps^2).
\end{equation}
We obtain
\begin{eqnarray}
\lefteqn{B_N(0,0,2,3,3,4)\ =\ \frac{2^{-6-2\eps}}{3(\Gamma(1+\eps))^2
  \Gamma(2-\eps)}\int_0^\infty K_{-\eps}(\xi)\left(K_{-1-\eps}(\xi)\right)^2
  K_{-2-\eps}(\xi)\xi^{7+2\eps}d\xi\ =}\nonumber\\
  &=&\frac{1+\eps}{192}\int_0^\infty K_0(\xi)\left(K_1(\xi)\right)^2K_2(\xi)
  \xi^7d\xi+\frac\eps{96}\int_0^\infty K_0(\xi)\left(K_1(\xi)\right)^2K_2(\xi)
  \xi^6d\xi\,+\qquad\qquad\nonumber\\&&
  +\frac\eps{96}\int_0^\infty K_0(\xi)\left(K_1(\xi)\right)^3\xi^6d\xi
  +\frac\eps{96}\int_0^\infty\left(K_0(\xi)\right)^2\left(K_1(\xi)\right)^2
  \xi^5d\xi\,+\nonumber\\&&+\frac\eps{96}\int_0^\infty K_0(\xi)
  \left(K_1(\xi)\right)^2K_2(\xi)\xi^7\ln(e^{\gamma_E}\xi/2)d\xi.
\end{eqnarray}

\subsection{The reduction procedure}
Especially in the last expression there are a lot of different integrals
found which differ from the basis $L_4(r)$ and $L_4^l(r)$. However, the
integrands can be reduced to integrand in terms of $K_0(\xi)$ and $K_1(\xi)$
only by using the relation
\begin{equation}
K_n(\xi)=2\frac{n-1}\xi K_{n-1}(x)+K_{n-2}(x).
\end{equation}
After the first step, namely the expansion of Bessel functions for
non-integer indices, the above relation establishes the second step in our
reduction procedure. Finally, we use
\begin{eqnarray}
\frac{d}{d\xi}K_0(\xi)&=&-K_1(\xi)\qquad\mbox{and}\nonumber\\
\frac{d}{d\xi}K_1(\xi)&=&-\frac12\left(K_0(\xi)+K_2(\xi)\right)
  \ =\ -K_0(\xi)-\frac1\xi K_1(\xi)
\end{eqnarray}
to perform the third and last step, for instance
\begin{eqnarray}
L_4^{(1)}(r)&=&\int_0^\infty\left(K_0(\xi)\right)^3K_1(\xi)\xi^rd\xi
  \ =\ -\int_0^\infty\left(K_0(\xi)\right)^3\frac{dK_0(\xi)}{d\xi}\xi^rd\xi
  \ =\nonumber\\
  &=&-\Big[K_0(\xi)\left(K_0(\xi)\right)^3\xi^r\Big]_0^\infty
  +\int_0^\infty K_0(\xi)\frac{d}{d\xi}\left(K_0(\xi)\right)^3\xi^rd\xi
  \ =\nonumber\\
  &=&3\int_0^\infty K_0(\xi)\left(K_0(\xi)\right)^2\frac{dK_0(\xi)}{d\xi}
  \xi^rd\xi+r\int_0^\infty K_0(\xi)\left(K_0(\xi)\right)^3\xi^{r-1}d\xi
  \ =\nonumber\\
  &=&-3\int_0^\infty\left(K_0(\xi)\right)^3K_1(\xi)\xi^rd\xi
  +r\int_0^\infty\left(K_0(\xi)\right)^4\xi^{r-1}d\xi
\end{eqnarray}
and therefore
\begin{equation}
L_4^{(1)}(r)=\int_0^\infty\left(K_0(\xi)\right)^3K_1(\xi)\xi^rd\xi
  =\frac r4\int_0^\infty\left(K_0(\xi)\right)^4\xi^{r-1}d\xi
  =\frac r4L_4(r-1).
\end{equation}
The complete set of reduction formulas is given by
\begin{eqnarray}
L_4^{(4)}(r)&=&(r-3)L_4^{(3)}(r-1)-3L_4^{(2)}(r),\nonumber\\[3pt]
L_4^{(3)}(r)&=&\frac12\left((r-2)L_4^{(2)}(r-1)-2L_4^{(1)}(r)\right),
  \nonumber\\
L_4^{(2)}(r)&=&\frac13\left((r-1)L_4^{(1)}(r-1)-L_4(r)\right),\nonumber\\
L_4^{(1)}(r)&=&\frac14rL_4(r-1),\\[7pt]
L_4^{l(4)}(r)&=&(r-3)L_4^{l(3)}(r-1)+L_4^{(3)}(r-1)-3L_4^{l(2)}(r),
  \nonumber\\[3pt]
L_4^{l(3)}(r)&=&\frac12\left((r-2)L_4^{l(2)}(r-1)+L_4^{(2)}(r-1)
  -2L_4^{l(1)}(r)\right),\nonumber\\
L_4^{l(2)}(r)&=&\frac13\left((r-1)L_4^{l(1)}(r-1)+L_4^{(1)}(r-1)
  -L_4^l(r)\right),\nonumber\\
L_4^{l(1)}(r)&=&\frac14\left(rL_4^l(r-1)+L_4(r-1)\right),
\end{eqnarray}
in general
\begin{eqnarray}\label{genred}
L_n^{(m)}(r)&=&\frac1{n-m+1}\left((r-m+1)L_n^{(m-1)}(r-1)
  -(m-1)L_n^{(m-2)}(r)\right),\\
L_n^{l(m)}(r)&=&\frac1{n-m+1}\left((r-m+1)L_n^{l(m-1)}(r-1)
  +L_n^{(m-1)}(r-1)-(m-1)L_n^{l(m-2)}(r)\right)\nonumber
\end{eqnarray}
and is coded in MATHEMATICA in order to automatically reduce to the master
integrals~\cite{Groote:1999cx}. The results for the examples after executing
the second and third step of the recursion reads
\begin{eqnarray}
B_N(0,0,2,2,2,2)&=&2(1+\eps)L_4(3)+4\eps L_4^l(3)+O(\eps^2),\nonumber\\[7pt]
B_N(0,0,2,2,2,1)&=&2L_4(1)+4\eps L_4^l(1)+O(\eps^2),\nonumber\\[3pt]
B_N(0,0,2,3,3,4)&=&\frac1{36}L_4(3)-\frac1{144}L_4(5)
  -\frac1{576}L_4(7)+O(\eps).
\end{eqnarray}
Comparing with the results obtained by using the RECURSOR
package~\cite{Broadhurst:1991fi},
\begin{eqnarray}
B_N(0,0,2,2,2,2)&=&-\frac38+\frac7{16}\zeta(3)+O(\eps),\nonumber\\
B_N(0,0,2,2,2,1)&=&\frac74\zeta(3)+O(\eps),\nonumber\\
B_N(0,0,2,3,3,4)&=&\frac1{576}+O(\eps)
\end{eqnarray}
(the order $\eps$ is omitted here) we can adjust the master integrals
\begin{equation}
L_4(3)=-\frac3{16}+\frac7{32}\zeta(3),\qquad
L_4(1)=\frac78\zeta(3),\qquad
16L_4(3)-2L_4(5)-L_4(7)=1.
\end{equation}
Especially the last relation is interesting because it means that
\begin{equation}
\int_0^\infty K_0(\xi)\left(K_1(\xi)\right)^2K_2(\xi)\xi^7d\xi=\frac13.
\end{equation}
This surprizing identity has been checked numerically as well. Note that
only $L_4(r)$ for odd values of $r$ are needed for the leading order
contribution. However, before concentrating on the master integrals
themselves, let me add some remarks on the efficiency of the reduction.

\subsection{The efficiency of the reduction}
Taking $N$ to be the maximal total power of denominator factors, the
calculation of a three-loop integral like in the previous examples would
create and deserve a field of $N^3$ integrals which are recursively defined
by each other using recurrence relations~\cite{Broadhurst:1991fi}. In contrast
to this, the reduction procedure in our case needs $N$ steps to reduce the
result to a basis of $2[N/2]-5$ basic elements $L_4(r)$ and $L_4^l(r)$.
Therefore, the ``costs'' of our method are of the order $N^2$ which
considerably reduces the time consumption in a computer evaluation.

\subsection{Values for the master integrals}
By comparing the results for different sun-rise type diagrams obtained by
using our reduction method~\cite{Groote:1999cx} and the procedure RECURSOR
in~\cite{Broadhurst:1991fi} we can find values for the master integrals.
The first results read
\begin{equation}
L_4(1)=\frac78\zeta(3),\qquad
L_4(3)=\frac7{32}\zeta(3)-\frac3{16},\qquad
L_4(5)=\frac{49}{128}\zeta(3)-\frac{27}{64},\ \ldots
\end{equation}
In general we found that the (odd-valued) master integrals are given by
expressions of the kind $L_4(2r+1)=A_r\zeta(3)-B_r$. Of course we might ask
whether there is a general way to calculate these master integrals. At least
we can execute the calculation for the master $L_4(1)$. This master integral
is (by a fresh view, i.e.\ not considering where the master integral comes
from) given by the three-loop sunrise-type bubble in two space-time dimensions
(i.e.\ $\lambda=0$),
\begin{equation}
\tilde\Pi(0)=\frac{2\pi^{\lambda+1}}{\Gamma(\lambda+1)}
  \int_0^\infty\pfrac{(mx)^\lambda K_\lambda(mx)}{(2\pi)^{\lambda+1}
  x^{2\lambda}}^4x^{2\lambda+1}dx=\frac1{(2\pi)^3m^2}\int_0^\infty
  \left(K_0(\xi)\right)^4\xi\,d\xi.
\end{equation}
$\tilde\Pi(0)$ iself can be written as the two-dimensional integral over the
product of two two-line correlators of the kind
\begin{equation}
\tilde\Pi_2(p)=\frac1{2\pi\sqrt{p^2}\sqrt{p^2+4m^2}}
  \ln\pfrac{\sqrt{p^2+4m^2}+\sqrt{p^2}}{\sqrt{p^2+4m^2}-\sqrt{p^2}}
  =\frac1{4\pi m^2}\frac\tau{\sinh\tau}
\end{equation}
(the last identity for substituting $p=2m\sinh(\tau/2)$), so that
\begin{eqnarray}
L_4(1)=(2\pi)^3m^2\int\frac{d^2p}{(2\pi)^2}\left(\tilde\Pi_2(p)\right)^2
  =\frac14\int_0^\infty\frac{\tau^2d\tau}{\sinh\tau}=\frac78\zeta(3)
\end{eqnarray}
where~\cite{Prudnikov}
\begin{equation}
\int_0^\infty\frac{\tau^{\alpha-1}d\tau}{\sinh\tau}
  =\frac{2^\alpha-1}{2^{\alpha-1}}\Gamma(\alpha)\zeta(\alpha)
\end{equation}
has been used. The expression for general (odd) values for $r$ reads
\begin{equation}
L_4(2r+1)=2\pi m^2\int\tilde\Pi(p)\left(-m^2\dalembertian_p\right)^r
  \tilde\Pi_2(p)d^2p
\end{equation}
where $\dalembertian_p=\partial^2/\partial p_\mu\partial p^\mu$ is the
two-dimensional d'Alembert operator in (Euclidean) momentum space. It is still
an open question whether this integral can be calculated to obtain the simple
expression containing rationals and $\zeta(3)$. However, a closer look at the
numerical value shows that, apart from the chance to obtain an analytical
result, the attempt might not be worth the effort. Indeed, looking at the
master integrals for higher and higher values of $r$ we find that the
contributions $A_r\zeta(3)$ and $B_r$ increase while the difference leads to
a significant cancellation, for $I_4(11)$ for instance up to three significant
figures. Instead of trying to calculate the analytical expression, we
therefore look for an asymptotic expansion of the integral at large values of
$r$,
\begin{equation}
L_4(2r+1)=\frac{\pi^2\Gamma(2r)}{4^{2r+1}}
  \left(1-\frac1{r-1/2}+O(1/r^2)\right).
\end{equation}
We can introduce a new parameter $\kappa$ which accounts for higher order
terms in the asymptotic expansion. The expression
\begin{equation}
L_4(2r+1)=\frac{\pi^2\Gamma(2r)}{4^{2r+1}}\left(1+\frac1{r+\kappa}\right)
\end{equation}
with $\kappa=0.97$ gives numerical results with an accuracy better than $1\%$
for all $r\ge 1$, while for $r>3$ the relative accuracy is better than
$10^{-3}$. A similar expression can be found for the logarithmic masters,
\begin{equation}
L_4^l(2r+1)=\frac{\pi\Gamma(2r)}{4^{2r+1}}\left(\Psi(2r)+\gamma_E-3\ln2\right)
  \left(1-\frac1{r+\kappa_l}\right)
\end{equation}
where $\Psi(z)=\Gamma'(z)/\Gamma(z)$ is the polygamma function or logarithmic
dericative of the gamma function. For $\kappa_l=1.17$ the accuracy for $r>3$
is better than $1\%$, for $r>5$ better than $10^{-3}$.

\subsection{Back to the genuine sunset}
As it is already visible in Eq.~(\ref{genred}), the recurrence relations just
discussed for $n=4$ massive internal lines are given for other numbers of
internal lines as well with no costs. We could for instance return to the
genuine sunset with $n=3$. In this case the master integral $L_3(1)$ is given
by a two-loop sunset-type bubble in two space-time dimensions,
\begin{equation}
\tilde\Pi(0)=\frac{2\pi^{\lambda+1}}{\Gamma(\lambda+1)}
  \int_0^\infty\pfrac{(mx)^\lambda K_\lambda(mx)}{(2\pi)^{\lambda+1}
  x^{2\lambda}}^3x^{2\lambda+1}dx=\frac1{(2\pi)^2m^2}
  \int_0^\infty\left(K_0(\xi)\right)^3\xi\,d\xi
\end{equation}
and, therefore ($t=e^{-\tau}$),
\begin{equation}
L_3(1;M/m)=m^2\int\frac{\tilde\Pi_2(p)}{p^2+M^2}d^2p=\int_0^\infty
  \frac{m^2\tau\,d\tau}{4m^2\sinh^2(\tau/2)+M^2}
  =-\int_0^1\frac{\ln t\,dt}{1-2\gamma t+t^2}
\end{equation}
where the mass $M$ of the third line is kept different from the other two
masses $m$ for a while. By differentiating with respect to $M$, any positive
power of the propagator (and/or power in $x$ in configuration space) can be
obtained -- therefore, all master integrals. Because of $\gamma=1-M^2/2m^2$,
the denominator has two roots $t_{1,2}=\gamma\pm\sqrt{\gamma^2-1}$ with
$t_1t_2=1$, and we obtain
\begin{equation}
L_3(1;M/m)=-\int_0^1\frac{\ln t\,dt}{(t-t_1)(t-t_2)}
  =-\frac{\Li_2(1/t_1)-\Li_2(1/t_2)}{t_1-t_2}
  =\frac{\Li_2(t_1)-\Li_2(t_2)}{t_1-t_2}
\end{equation}
where $\Li_2(z)$ is the dilogarithm function. The differentiation with respect
to $M$ is now straightforward and can be performed with a symbolic manipulation
program. The pseudo-threshold case $M=2m$ is a degenerate case, we obtain
\begin{equation}
L_3(1;2)=-\int_0^1\frac{\ln t\,dt}{(1+t)^2}=\ln 2
\end{equation}
while for $M=m$ we obtain $\gamma=1/2$, $t_{1,2}=\exp(\pm i\pi/3)$, and
therefore
\begin{equation}
L_3(1)=L_3(1;1)=-\int_0^1\frac{\ln t\,dt}{1-t+t^2}=\frac2{\sqrt3}
  {\rm Im\,}\Li_2(e^{i\pi/3})=\frac2{\sqrt3}{\rm Cl}_2\pfrac\pi3,
\end{equation}
a result in terms of Clausen's dilogarithms.

\subsection{Generalization to the spectacle topology}
\stepcounter{figure}%
\parbox[b]{6.5truecm}{\epsfig{figure=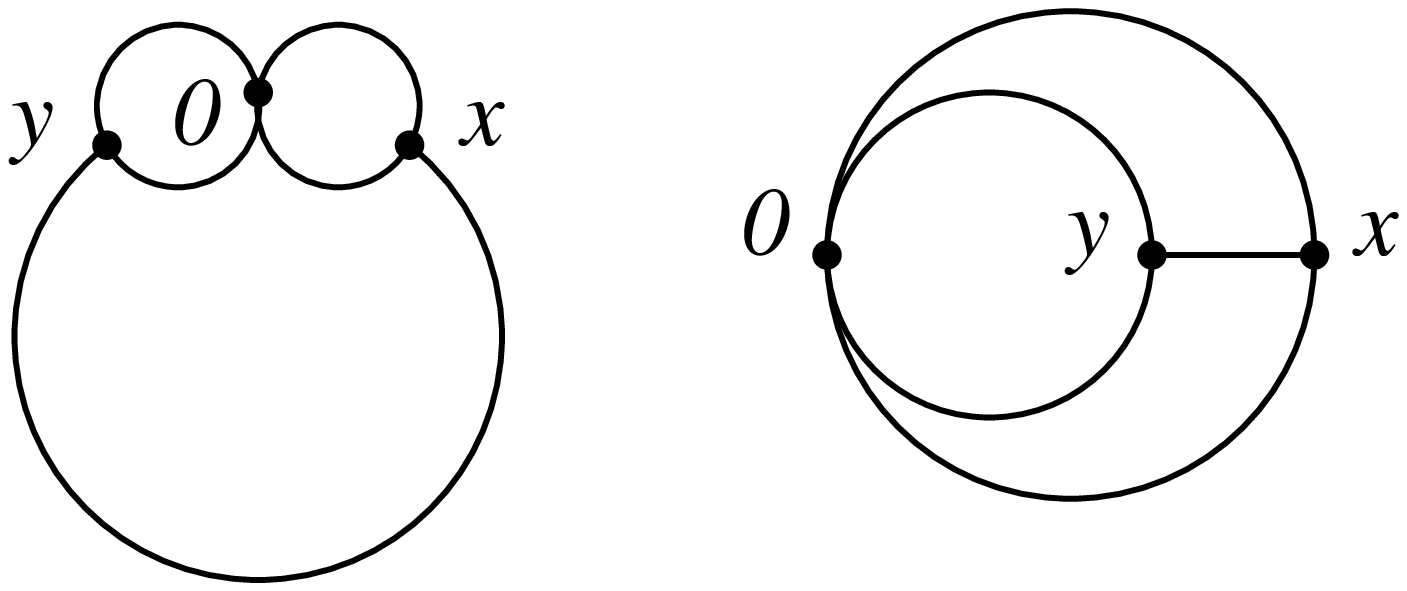, scale=0.4}\\
Figure~\arabic{figure}: ``spectacle+propagator'' representation, also called
``spectacle'' topology diagram, in two different forms. The configuration
space points $0$, $x$ and $y$ are indicated.
\vspace{7pt}}\hfill
\parbox[b]{9truecm}{To finish today's lecture, we can look a bit beyond the
rim of the plate of sunrise-type diagrams. Actually, we can generalize the
configuration space technique to the so-called spectacle topology, as shown in
Fig.~\arabic{figure} in two different representations. The configuration space
expression of a spectacle topology diagram (again with a different mass $M$ on
the frame) written in a form suitable for the actual purpose is given by}
\begin{equation}
S(M)=\int D(x-y,M)D(x,m)^2D(y,m)^2d^Dx\,d^Dy.
\end{equation}
The key relation for a significant simplification of the configuration space
integral with spectacle topology is given by a relation obtained from the
addition theorem of Bessel functions~\cite{Gradshteyn}. I will not go into
detail here, only mention the result~\cite{Groote:1999cx}
\begin{equation}
\int\frac{K_\lambda(|r-\rho|)}{(2\pi)^{\lambda+1}|r-\rho|^\lambda}d\Omega_\rho
  =\frac{I_\lambda(\rho)}{\rho^\lambda}\frac{K_\lambda(r)}{r^\lambda},
  \qquad r>\rho
\end{equation}
where $I_\lambda(z)$ is the modified Bessel function of the first kind. This
result allows one to write
\begin{eqnarray}
S(M)&=&\int\frac{M^{2\lambda}}{(2\pi)^{\lambda+1}}
  \frac{K_\lambda(M|x-y|)}{(M|x-y|)^\lambda}D(x,m)^2D(y,m)^2d^Dx\,d^Dy
  \ =\nonumber\\
  &=&\pfrac{2\pi^{\lambda+1}}{\Gamma(\lambda+1)}^2M^{2\lambda}
  \int_0^\infty\left(D(x,m)\right)^2x^{2\lambda+1}dx
  \int_0^\infty\left(D(y,m)\right)^2x^{2\lambda+1}dy\
  \times\nonumber\\&&\qquad\qquad\times\
  \left(\frac{K_\lambda(Mx)}{(Mx)^\lambda}\frac{I_\lambda(My)}{(My)^\lambda}
  \theta(x-y)+\frac{K_\lambda(My)}{(My)^\lambda}
  \frac{I_\lambda(Mx)}{(Mx)^\lambda}\theta(y-x)\right).
\end{eqnarray}
As an illustration we look at the two-dimensional space-time where
\begin{eqnarray}
S(M)&=&\frac1{(2\pi)^2}\int_0^\infty x\left(K_0(mx)\right)^2dx
  \int_0^\infty y\left(K_0(my)\right)^2dy\ \times\nonumber\\&&\qquad\times
  \left(K_0(Mx)I_0(My)\theta(x-y)+K_0(My)I_0(Mx)\theta(y-x)\right),
\end{eqnarray}
comparing this result with the momentum space representation
\begin{equation}
S(M)=\int\frac{d^2p}{(2\pi)^2}\frac{\left(\tilde\Pi_2(p)\right)^2}{p^2+M^2}
  =\frac{2\pi}{m^4}\int_0^1\frac{t\,\ln^2t\,dt}{(1-t^2)(1-2\gamma t+t^2)}
\end{equation}
where the same chai of substitutions have been used. We obtain
\begin{equation}
S(M)=\frac{f(t_1)-f(t_2)}{t_1-t_2},\qquad
f(t)=\frac{4\pi t\Li_3(1/t)-(t+7)\zeta(3)}{m^4(t^2-1)}
\end{equation}
where $\Li_3(z)$ is the trilogarithm function. While for $M=m$ we obtain a
result including the Clausen trilogarithm ${\rm Cl}_3(2\pi/3)$, for the case
$M=2m$ the integral for $S(M=2m)$ simplifies as in the case of the sunrise
diagram, one obtains
\begin{equation}
S(2m)=\frac\pi{m^4}\left(\frac78\zeta(3)-\ln 2\right).
\end{equation}
In all these cases the sixth order roots of unity $\pm 1$, $\exp(\pm\pi/3)$,
and $\exp(\pm2\pi/3)$ play an important role~\cite{Broadhurst:1998rz}.
In~\cite{Groote:1999cx,Groote:1999cn} we have followed the lines given
in~\cite{Broadhurst:1998rz} and have constructed a shuffle algebra for the
occuring integrals which finally lead to the transcendental numbers which
normally occur in these calculations. At this point I will not go into detail.
Instead I close the lecture by mentioning that the numerical comparison
between the result obtained in the momentum space approach with the result
within the configuration space technique show conicidence. Tomorrow we will
deal with the consideration of sunrise-type diagrams close to threshold.

\newpage

\section{Expansions close to threshold}
From the physical point of view the interesting part of our analysis of
sunrise-type diagrams is the construction of the spectral decomposition of the
diagram. Using the configuration space technique, the spectral density can be
constructed directly from the correlator function in configuration space. The
technique introduced in~\cite{Groote:1998ic} is based on an integral
transform. Starting with the dispersion relation in (Euclidean) momentum space,
\begin{equation}
\tilde\Pi(q)=\int_0^\infty\frac{\rho(s)ds}{q^2+s}
  =\int_0^\infty\rho(s)\tilde D(q,\sqrt s)ds,
\end{equation}
the corresponding relation in configuration space reads
\begin{equation}
\Pi(x)=\int_0^\infty\rho(s)D(x,\sqrt s)ds.
\end{equation}
This representation was used for sum rule applications
in~\cite{Pivovarov:ij,Chetyrkin:yr} where the spectral density for the
two-loop sunrise diagram was found in two-dimensional
space-time~\cite{Pivovarov:jm}. With the explicit form of the propagator in
configuration space given by
\begin{equation}
D(x,\sqrt s)=\frac{(\sqrt s/x)^\lambda}{(2\pi)^{\lambda+1}}K_\lambda(\sqrt sx),
\end{equation}
the representation turns into a particular example of the Hankel transform,
namely the $K$-transform~\cite{Meijer,Erdelyi}
\begin{equation}
g(y)=\int_0^\infty f(x)K_\nu(xy)\sqrt{xy}\,dx.
\end{equation}
The inverse of this transform is known to be given by
\begin{equation}
f(x)=\frac1{i\pi}\int_{c-i\infty}^{c+i\infty}g(y)I_\nu(xy)\sqrt{xy}\,dy
\end{equation}
where the vertical contour in the complex plane is placed to the right of the
right-most singularity of the function $g(y)$~\cite{Erdelyi}. Translating this
inverse $K$-transform to our situation which relates $\rho(s)$ and $\Pi(x)$,
we obtain
\begin{equation}\label{invKtrans}
\sqrt s^\lambda\rho(s)=-(2\pi)^\lambda i\int_{c-i\infty}^{c+i\infty}
  \Pi(x)x^{\lambda+1}I_\lambda(\sqrt sx)dx.
\end{equation}
The inverse transform given in Eq.~(\ref{invKtrans}) solves the problem of
determining the spectral density of sunrise-type diagrams by reducing it to
the computation of a one-dimensional integral for the general class of
sunrise-type diagrams with any number of internal lines and different masses.
Because the contour can bypass the area of small values of $x$, the integral
is finite. Using this expression, the spectral density is known numerically
as a function of~$s$. Eq.~(\ref{invKtrans}) will be used in the following to
compare the exact result with different types of expansions found in the
literature, in order to test their convergence behaviour.

\subsection{Considerations in Minkowskian domain}
The threshold region of a sunrise-type diagram is determined by the condition
$q^2+M^2\cong 0$, where $q$ is the Euclidean momentum and $M=\sum_im_i$ is the
threshold value for the spectral density. We introduce the Minkowskian momentum
$p$ defined by $p^2=-q^2$ which is an analytic continuation to the physical
cut. Operationally, this analytic continuation can be performed by replacing
$q\rightarrow ip$. To analyze the region near the threshold we use the
parameter $\Delta=M-p$ which takes complex values, while in phenomenological
applications the parameter $E=-\Delta=p-M$ is more convenient. The spectral
density will be written as a function of $E$ in the following,
$\tilde\rho(E)=\rho((M+E)^2)$. The analytic continuation of the Fourier
transform $\tilde\Pi(q)$ to the Minkowskian domain has the form
\begin{equation}
\tilde\Pi_M(p)=2\pi^{\lambda+1}\int_0^\infty\pfrac{ipx}2^{-\lambda}
  J_\lambda(ipx)\Pi(x)x^{2\lambda+1}dx
\end{equation}
(as we work in Minkowskian domain only, the index $M$ will be dropped in the
following). 

\subsection{Large $x$ behaviour of the weight}
For the threshold expansion we have to analyze the large $x$ behaviour of the
integrand. It is this region that saturates the integral in the limit
$p\rightarrow M$ or, equivalently, $E\rightarrow 0$. It is convenient to
perform the analysis in a basis where the integrand has a simple large $x$
behaviour. The most important part of the integrand is the Bessel function
$J_\lambda(ipx)$ which, however, contains both rising and falling branches at
large $x$. It resembles the situation with the elementary trigonometric
function $\cos(z)$ to which the Bessel function $J_\lambda(z)$ is rather close
in a certain sense, as we will see. $\cos(z)$ is a linear combination of
exponentials,
\begin{equation}
\cos(z)=\frac12\left(e^{iz}+e^{-iz}\right).
\end{equation}
In the same manner, the Bessel function $J_\lambda(z)$ can be written as a sum
of two Hankel functions $H_\lambda^\pm(z)=J_\lambda(z)\pm iY_\lambda(z)$,
\begin{equation}
J_\lambda(z)=\frac12\Big(H_\lambda^+(z)+H_\lambda^-(z)\Big),
\end{equation}
and for pure imaginary argument the Hankel functions show simple asymptotic
behaviour
\begin{equation}
H_\lambda^\pm(iz)\sim z^{-1/2}e^{\pm z}.
\end{equation}
Accordingly, we split up $\tilde\Pi(p)$ into
$\tilde\Pi(p)=\tilde\Pi^+(p)+\tilde\Pi^-(p)$ with
\begin{equation}
\tilde\Pi^\pm(p)=\pi^{\lambda+1}\int_0^\infty\pfrac{ipx}2^{-\lambda}
  H_\lambda^\pm(ipx)\Pi(x)x^{2\lambda+1}dx.
\end{equation}
The two parts $\tilde\Pi^\pm(p)$ of the polarization function $\tilde\Pi(p)$
have completely different behaviour near threshold and are analyzed
independently in the following.

\subsection{The polarization function $\tilde\Pi^+(p)$}
The behaviour of $\tilde\Pi^+(p)$ for large $x$ is given by the asymptotic
form of the functions which I simply write up to the leading terms as
\begin{equation}
H^+(ipx)=\sqrt{\frac2{i\pi px}}e^{-px}(1+O(x^{-1})),\qquad
K(mx)=\sqrt{\frac\pi{2mx}}e^{-mx}(1+O(x^{-1})).
\end{equation}
We conclude that the large $x$ range of the integral (above a reasonably large
cutoff parameter $\Lambda$) has the general form
\begin{equation}
\tilde\Pi_\Lambda^+(M-\Delta)\sim\int_\Lambda^\infty x^{-a}e^{-(2M-\Delta)x}dx,
\qquad a=(n-1)(\lambda+1/2).
\end{equation}
The right hand side is an analytic function in $\Delta$ in the vicinity of
$\Delta=0$. It exhibits no cut or other singularities near the threshold and
therefore does not contribute to the spectral density. I could therefore go
ahead to the second contribution $\tilde\Pi^-(p)$. However, I want to do a
bit more in order to fill $\tilde\Pi^+(p)$ with ``life'' (i.e., meaning). In
using the relation
\stepcounter{figure}%
\parbox[b]{6.5truecm}{\hfil\epsfig{figure=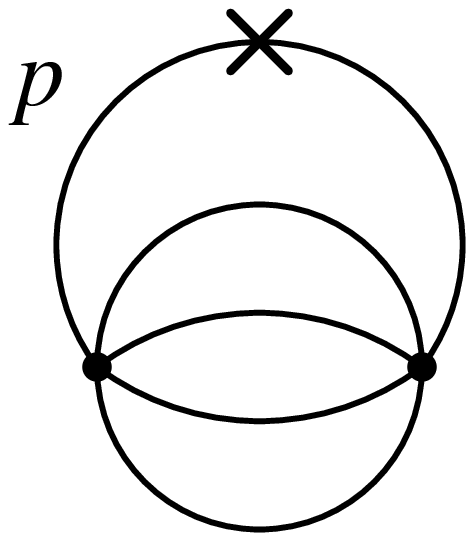, scale=0.4}\\
Figure~\arabic{figure}: Representation of the regular part
$\tilde\Pi^+(-p^2)$ as vacuum bubble with added line. The cross denotes an
arbitrary number of derivatives of the specified line.\vspace{7pt}}\hfill
\parbox[b]{9truecm}{\begin{equation}
K_\lambda(z)=\frac{\pi i}2e^{i\lambda\pi/2}H_\lambda^+(iz)
\end{equation}
between Bessel functions fo different kind I can replace
the Hankel function $H_\lambda^+(ipx)$ by the Bessel function
$K_\lambda(px)$. But since the propagator of a massive particle (massive line
in the diagram) is given by the Bessel function $K_\lambda(mx)$ up to a power
$x$, the weight function behaves like a propagator of an additional line with
``mass'' $p$. The explicit expression is given by (cf.\ Fig.~\arabic{figure})}
\begin{equation}
\tilde\Pi^+(p)=\frac{(-2\pi i)^{2\lambda+1}}{(p^2)^\lambda}
  \int_0^\infty\Pi_+(x)x^{2\lambda+1}dx,\qquad
\Pi_+(x)=\Pi(x)D(x,p).
\end{equation}

\subsection{The polarization function $\tilde\Pi^-(p)$}
In contrast to the previous case, the integrand of the part $\tilde\Pi^-(p)$
contains $H^-(ipx)$ which behaves like a rising exponential function at large
$x$,
\begin{equation}
H^-(ipx)\sim x^{-1/2}e^{px}.
\end{equation}
Therefore, the integral has the large $x$ behaviour
\begin{equation}
\tilde\Pi_\Lambda^-(M-\Delta)\sim\int_\Lambda^\infty x^{-a}e^{-\Delta x}dx.
\end{equation}
The exponential factor $e^{-\Delta x}$ which establishes an effective inverse
proportionality between $\Delta$ and $x$ is the reason for the importance of
the large $x$ behaviour as being essential for the near threshold expansion of
the spectral density. However, for $\Delta<0$ the integral diverges at the
upper limit, leading to non-analyticity at the threshold $\Delta=0$. In the
complex $\Delta$ plane with a cut along the negative real axis the integral is
analytic. This cut corresponds to the physical positive energy cut. The
discontinuity across the cut gives rise to the non-vanishing spectral density
of the contribution $\tilde\Pi^-(p)$.

In order to obtain an expansion for the spectral density near the threshold in
an analytical form we make use of the asymptotic series expansion for the
function $\Pi(x)$ that crucially simplifies the integrands but still preserves
the singular structure of the integral in terms of the variable $\Delta$.
Because of the already mentioned asymptotic expansion
\begin{equation}
K_\lambda^{\rm as}(mx)=\sqrt{\frac\pi{2mx}}e^{-mx}\left(1+O(x^{-1})\right),
\end{equation}
the asymptotic expansion of the function $\Pi(x)$ consists of an
exponential factor $e^{-Mx}$ and an inverse power series in $x$ up to an order
$\tilde N$ closely related to $N$. It is this asymptotic expansion that
determines the singularity structure of the integral. We write the whole
integral in the form of the sum of two terms again,
\begin{eqnarray}\label{pidias}
\tilde\Pi^-(p)&=&\pi^{\lambda+1}\int\pfrac{ipx}2^{-\lambda}H_\lambda^-(ipx)
  \left(\Pi(x)-\Pi^{as}_N(x)\right)x^{2\lambda+1+2\eps}dx\,+\nonumber\\&&
  +\pi^{\lambda+1}\int\pfrac{ipx}2^{-\lambda}H_\lambda^-(ipx)
  \Pi^{as}_N(x)x^{2\lambda+1+2\eps}dx=\tilde\Pi^{di}(p)+\tilde\Pi^{as}(p).
\end{eqnarray}
The integrand of the difference term $\tilde\Pi^{di}(p)$ behaves as
$1/x^{\tilde N}$ at large $x$ while the integrand of the second term
accumulates all lower powers of the large $x$ expansion. We will again
consider the two parts in turn. However, a comment on the regularization is
in order here.

\subsection{Comment on the regularization used}
Being interested only in the large $x$ behaviour, we introduced a cutoff
$\Lambda$ which necessary at least at this point because the asymptotic
expansions are not defined for small values of $x$ and lead to divergences.
However, from the practical point of view the calculation of the regularized
integrals with an explicit cutoff is inconvenient. Instead, the standar way to
cope with such a situation is to use dimensional regularization. Note that
dimensional regularization does not necessarily regularize all divergences in
this case (in contrast to the standard case of ultraviolett divergences) but
nevertheless is sufficient for our purposes. Therefore, we use a parameter
$\eps$ independent of $\lambda$ to regularize the divergences at small~$x$.

\subsection{The difference part $\tilde\Pi^{di}(p)$}
In $\tilde\Pi^{di}(p)$ the subtracted asymptotic series to order $N$ cancels
the inverse power behaviour of the integrand to this degree $N$. Therefore,
the integrand decreases sufficently fast for large values of $x$ and the
integral even converges for $\Delta=0$. The part $\tilde\Pi^{di}(p)$ is
regular and gives no contribution to the spectral density up to the order
$\Delta^N$.

\subsection{The asymptotic part $\tilde\Pi^{as}(p)$}
The integral $\tilde\Pi^{as}(p)$ is still rather complicated to compute. But
we can go a step further in its analytical evaluation. Indeed, since the
singular behaviour of $\tilde\Pi^{as}(p)$ is determined by the behaviour at
large $x$, we can replace the first factor, i.e.\ the Hankel function, in
the large $x$ region by its asymptotic expansion up to some order $N$,
\begin{equation}
H_{\lambda,N}^{-as}(iz)=\sqrt{\frac2{\pi z}}e^{z+i\lambda\pi/2}
  \left[\sum_{n=0}^{N-1}\frac{(-1)^n(\lambda,n)}{(2z)^n}
  +\theta\frac{(-1)^N(\lambda,N)}{(2z)^N}\right]
\end{equation}
with
\begin{equation}
\theta\in[0,1],\qquad
(\lambda,n):=\frac{\Gamma(\lambda+n-1/2)}{n!\Gamma(\lambda-n-1/2)}
\end{equation}
to obtain the double asymptotic representation
\begin{equation}
\tilde\Pi^{das}(p)=\pi^{\lambda+1}\int_0^\infty\pfrac{ipx}2^{-\lambda}
  H_{\lambda,N}^{-as}(ipx)\Pi_N^{as}(x)x^{2\lambda+1+2\eps}dx.
\end{equation}
Both asymptotic expansions are straightforward and can be obtained from
standard handbooks on Bessel functions. We therefore arrive at our final
result: the integration necessary for evaluating the near threshold expansion
of the sunrise-type diagrams reduces to integrals of Euler's Gamma function
type, i.e.\ integrals containing exponentials and powers in the integrand.
Indeed, the result of the asymptotic expansion of the integrand is an
exponential function $e^{-\Delta x}$ times a power series in $1/x$, namely
\begin{equation}
x^{-a+2\eps}e^{-\Delta x}\sum_{j=0}^{N-1}\frac{A_j}{x^j}
\end{equation}
where $a=(n-1)(\lambda+1/2)$ has already been used earlier and the
coefficients $A_j$ are simple functions of the momentum $p$ and the masses
$m_i$. This expression can be integrated analytically using
\begin{equation}
\int_0^\infty x^{-a+2\eps}e^{-\Delta x}dx=\Gamma(1-a+2\eps)\Delta^{a-1-2\eps}.
\end{equation}
The result is
\begin{equation}\label{serint}
\tilde\Pi^{das}(M-\Delta)
  =\sum_{j=0}^{N-1}A_j\Gamma(1-a-j+2\eps)\Delta^{a+j-1-2\eps}.
\end{equation}
This expression is our final representation for the part of the polarization
function of a sunrise-type diagram necessary for the calculation of the
spectral density near the production threshold. Starting from this main result
in Eq.~(\ref{serint}), I discuss the general structure in some detail. In the
case where $a$ takes integer values, these coefficients result in
$1/\eps$-divergences for small values of $\eps$. The powers of $\Delta$ in
Eq.~(\ref{serint}) have to be expanded to first order in $\eps$ and give
\begin{equation}
\frac1{2\eps}\Delta^{2\eps}=\frac1{2\eps}+\ln\Delta+O(\eps).
\end{equation}
Because of
\begin{equation}
{\rm Disc\,}\ln(\Delta) \equiv \ln(-E-i0)-\ln(-E+i0)
=-2\pi i\theta(E)
\end{equation}
$\tilde\Pi^{das}(M-\Delta)$ in Eq.~(\ref{serint}) contributes to the spectral
density. For half-integer values of $a$ the power of $\Delta$ itself has a
cut even for $\eps=0$. The discontinuity is then given by
\begin{equation}
{\rm Disc\,}\sqrt{\Delta}=-2i\sqrt{E}\,\theta(E).
\end{equation}
Our method to construct a threshold expansion thus simply reduces to the
analytical calculation of the part $\tilde\Pi^{das}(p)$ which can be done
for arbitrary dimension and an arbitrary number of lines with different
masses. In the next subsections I work out some specific examples which
demonstrate both the simplicity and efficiency of our method.

\subsection{First example: equal mass ($n=3$) sunrise diagram}
\begin{figure}[t]\begin{center}
\epsfig{figure=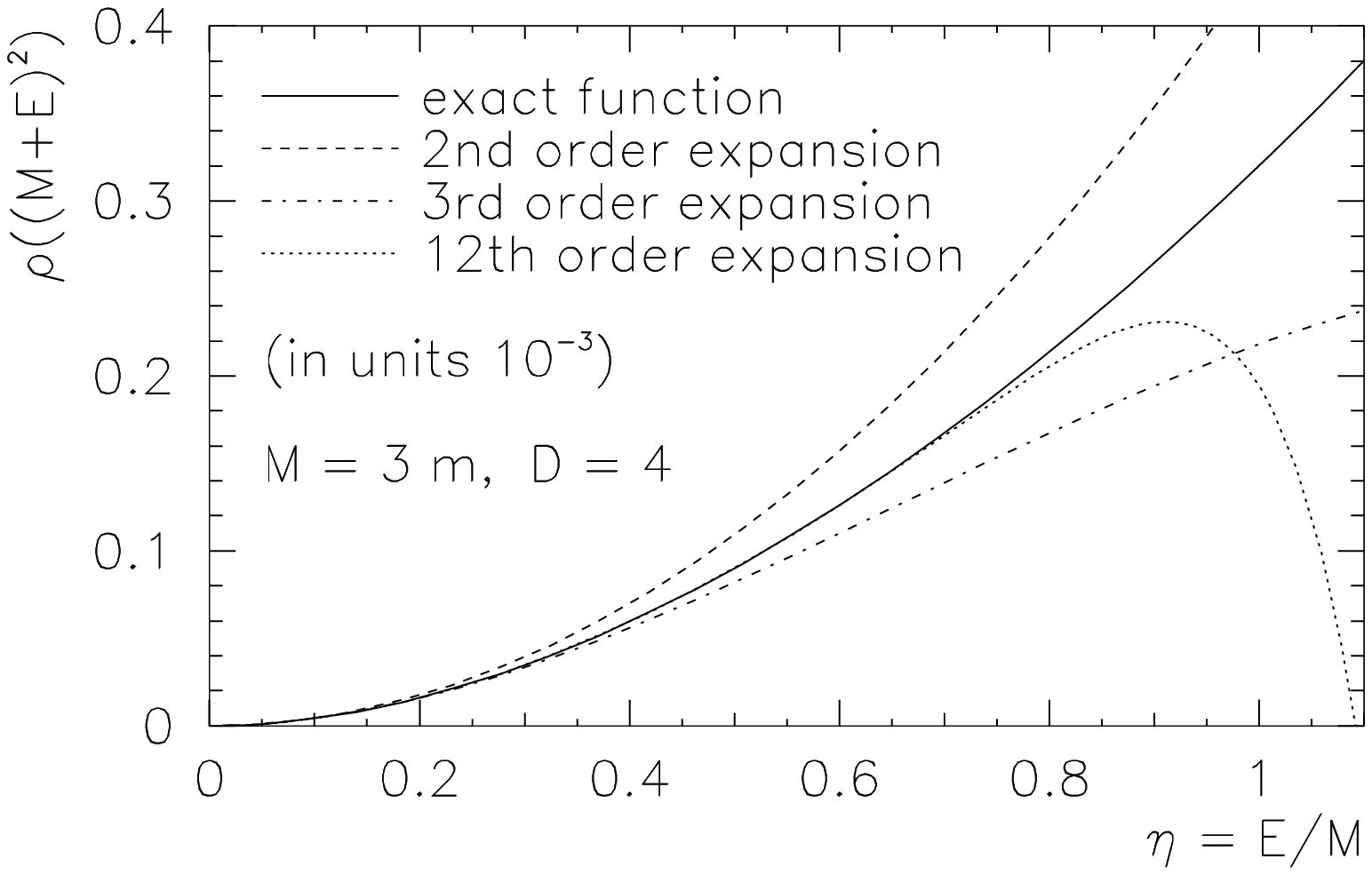, scale=0.8}\end{center}
\caption{\label{fig05}Various results for the spectral density for $n=3$ equal
masses in $D=4$ space-time dimensions in dependence on the threshold parameter
$E/M$. Shown are the exact solution obtained by using Eq.~(\ref{invKtrans})
(solid curve) and threshold expansions for different orders taken from
Eq.~(\ref{pidas430}) (dashed to dotted curves).}
\end{figure}

For the genuine sunrise diagram with equal masses (the last restriction for
reasons of representation only), the double asymptotic expansion reads
\begin{eqnarray}
\lefteqn{\pi^2\pfrac{ipx}2^{-1}H_{1,N}^{as}(px)\Pi_N^{as}(x)
  x^{3+2\eps}\ =\ \frac{m^{3/2}e^{(p-3m)x}}{(4\pi)^3p^{3/2}}x^{-3+2\eps}\
  \times}\nonumber\\&&\qquad\times\ \left\{1+\frac9{8mx}-\frac3{8px}
  +\frac9{128m^2x^2}-\frac{27}{64mpx^2}-\frac{15}{128p^2x^2}+O(x^{-3})\right\}.
\end{eqnarray}
The spectral density is obtained by performing the term-by-term integration of
the series and by evaluating the discontinuity across the cut along the
positive energy axis $E>0$. The result reads
\begin{eqnarray}\label{pidas430}
\lefteqn{\tilde\rho(E)\ =\ \frac{E^2}{384\pi^3\sqrt 3}
  \Bigg\{1-\frac12\eta+\frac7{16}\eta^2-\frac38\eta^3
  +\frac{39}{128}\eta^4-\frac{57}{256}\eta^5}\qquad\\&&
  +\frac{129}{1024}\eta^6-\frac3{256}\eta^7
  -\frac{4047}{32768}\eta^8+\frac{18603}{65536}\eta^9
  -\frac{248829}{524288}\eta^{10}+O(\eta^{11})\Bigg\}\nonumber
\end{eqnarray}
where the notation $\eta=E/M$, $M=3m$ is used~\cite{Groote:2000kz}. The
simplicity of this derivation is striking and easily reproduces the expansion
coefficients in~\cite{Berends:1997vk}. For different orders the result is
compared with the exact result from inverse $K$-transform
(Eq.~(\ref{invKtrans})) in Fig.~\ref{fig05}.

\begin{figure}[t]\begin{center}
\epsfig{figure=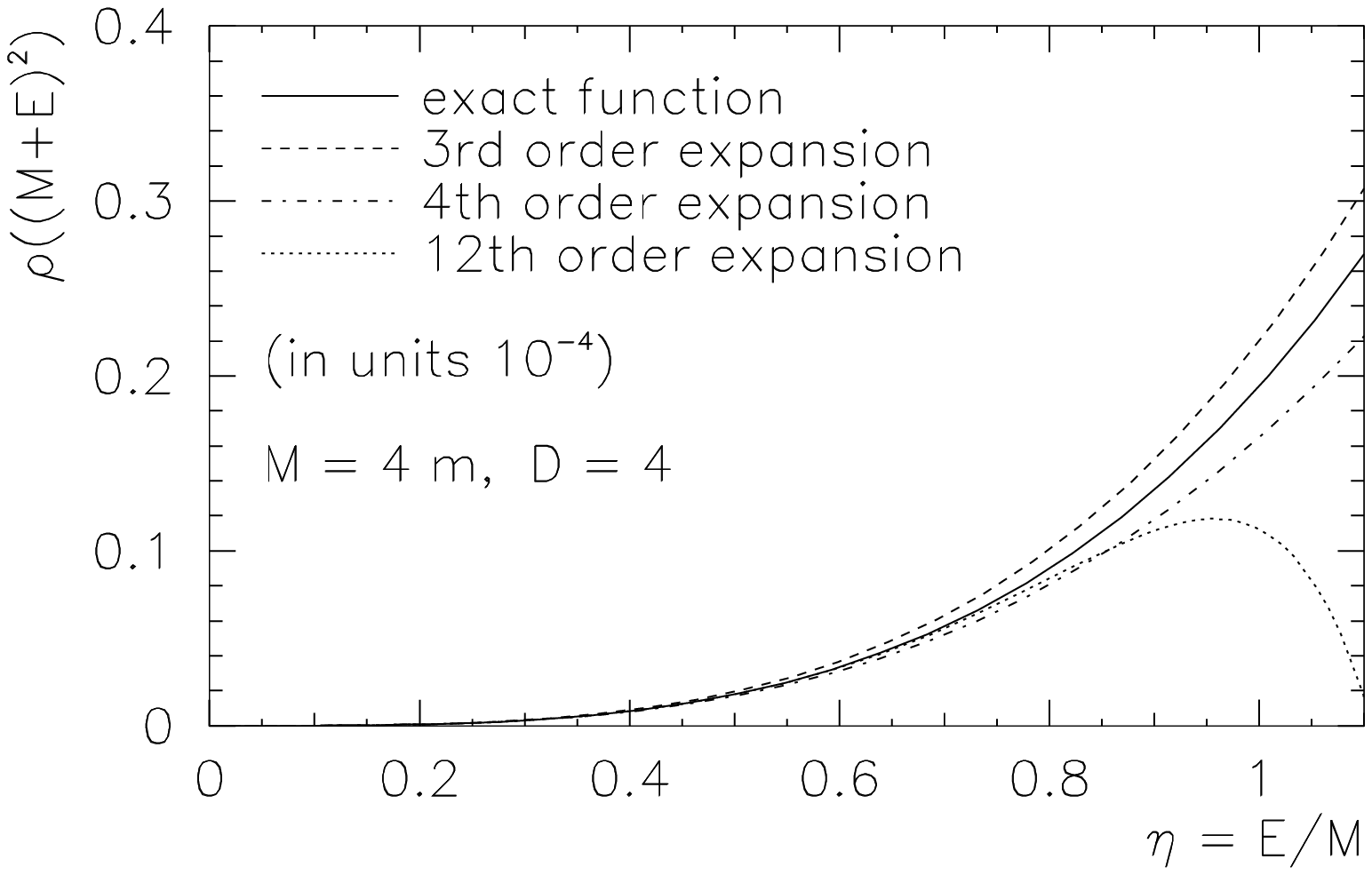, scale=0.8}\end{center}
\caption{\label{fig06}Various results for the spectral density for $n=4$ equal
masses in $D=4$ space-time dimensions in dependence on the threshold parameter
$E/M$. Shown are the exact solution obtained by using Eq.~(\ref{invKtrans})
(solid curve) and threshold expansions for different orders taken from
Eq.~(\ref{pidas440}) (dashed to dotted curves).}
\end{figure}

\subsection{Second example: equal mass $n=4$ sunrise-type diagram}
The sunrise-type diagram with four or more propagators cannot be easily done
by using the momentum space technique because it requires the multiloop
integration of entangled momenta. Within the configuration space technique
the generalization to any number of lines (or loops) is immediate by no effort.
For the spectral density of the equal mass $n=4$ sunrise-type diagram we obtain
(cf.\ Fig.~\ref{fig06})
\begin{eqnarray}\label{pidas440}
\lefteqn{\tilde\rho(E)\ =\ \frac{E^{7/2}M^{1/2}}{26880\pi^5\sqrt2}
\Bigg\{1-\frac14\eta+\frac{81}{352}\eta^2-\frac{2811}{18304}\eta^3
  +\frac{17581}{292864}\eta^4}\\&&\kern-16pt
  +\frac{1085791}{19914752}\eta^5-\frac{597243189}{3027042304}\eta^6
  +\frac{4581732455}{12108169216}\eta^7
  -\frac{496039631453}{810146594816}\eta^8+O(\eta^9)\Bigg\}\nonumber
\end{eqnarray}
where $\eta=E/M$ and $M=4m$ is the threshold value. One sees the difference
with the previous three-line case. In Eq.~(\ref{pidas440}) the cut represents
the square root branch while in the three-line case it was a logarithmic cut.
This underlines my earlier considerations.

\subsection{Strongly asymmetric mass arrangement}
As we have seen, the threshold expansion for equal (or close) masses breaks
down for $E\sim M=\sum m_i$. However, if the masses are not equal, the region
of the breakdown of the expansion is determined by the mass with the smallest
numerical value. Actually, one finds a breakdown at $E\sim 2m_0$ where $m_0$
is a mass much smaller than the others. To our best knowledge, this issue has
not been touched earlier. In the following I will describe a technique called
resummation of the smallest mass contribution~\cite{Groote:2000kz}. Instead of
the double asymptotic expansion we use the partial asymptotic expansion
\begin{equation}
\tilde\Pi^{pas}(p)=\pi^{\lambda+1}\int_0^\infty\pfrac{ipx}2^{-\lambda}
  H_{\lambda,N}^{-as}(ipx)\Pi_{m_0}^{as}(x)x^{2\lambda+1+2\eps}dx
\end{equation}
where the asymptotic expansions are substituted for all propagators except
for the one with the smallest mass $m_0$,
\begin{equation}
\Pi_{m_0}^{as}(x)=\Pi_{n-1}^{as}(x)D(m_0,x).
\end{equation}
Because of the knowledge of
\begin{eqnarray}
\lefteqn{\int_0^\infty x^{\mu-1}e^{-\tilde\alpha x}K_\nu(\beta x)dx
  \ =}\nonumber\\
  &=&\frac{\sqrt\pi(2\beta)^\nu}{(2\tilde\alpha)^{\mu+\nu}}
  \frac{\Gamma(\mu+\nu)\Gamma(\mu-\nu)}{\Gamma(\mu+1/2)}\
  {_2F_1}\left(\frac{\mu+\nu}2,\frac{\mu+\nu+1}2;\mu+\frac12;
  1-\frac{\beta^2}{\tilde\alpha^2}\right)
\end{eqnarray}
with $\tilde\alpha=\Delta-m_0$ and $\beta=m_0$, $\tilde\Pi^{pas}(p)$ is
expressible in terms of hypergeometric functions~\cite{AbramowitzStegun}.
The same is valid for the spectral density, because
\begin{eqnarray}\label{genuine}
\lefteqn{\frac1{2\pi i}\Disc\int_0^\infty x^{\mu-1}e^{\alpha x}
  K_\nu(\beta x)dx\ =}\nonumber\\
  &=&\frac{2^\mu(\alpha^2-\beta^2)^{1/2-\mu}}{\alpha^{1/2-\nu}\beta^\nu}
  \frac{\Gamma(3/2)}{\Gamma(3/2-\mu)}\ {_2F_1}\left(\frac{1-\mu-\nu}2,
  \frac{2-\mu-\nu}2;\frac32-\mu;1-\frac{\beta^2}{\alpha^2}\right)\qquad
\end{eqnarray}
where $\alpha=E+m_0$. For the case of an $n=2$ sunrise-type diagram with
masses $m$ and $m_0\ll m$ we compare the exact result in Eq.~(\ref{invKtrans})
for the spectral density with the pure expansion
near threshold
\begin{figure}[t]\begin{center}
\epsfig{figure=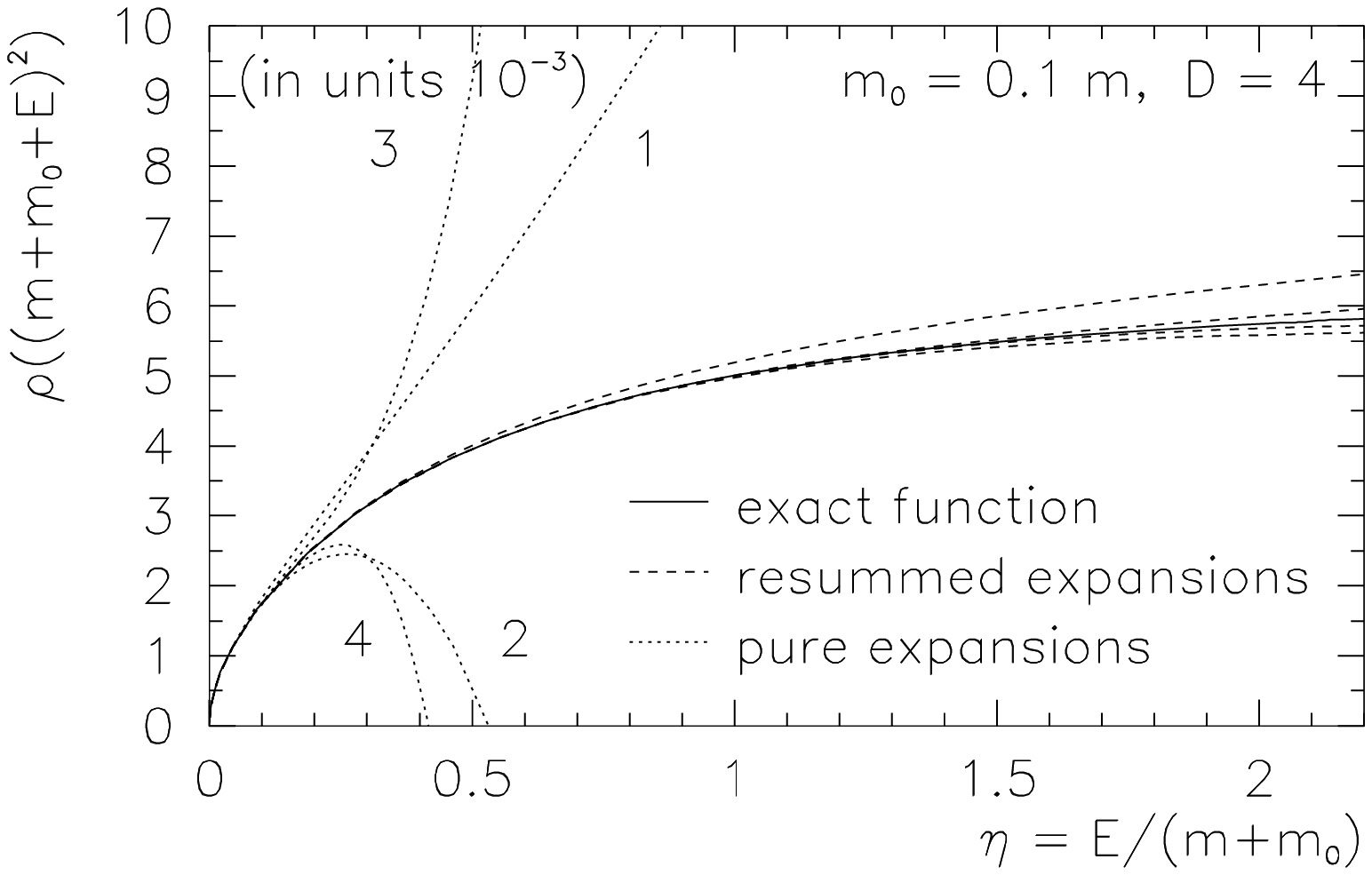, scale=0.8}\end{center}
\caption{\label{fig07}Various solutions for the spectral density for two
masses $m$ and $m_0\ll m$ and $D=4$ space-time dimensions. Shown are the exact
solution which is obtained by using Eq.~(\ref{invKtrans}) (solid curve), the
pure threshold expansions using Eq.~(\ref{pidas42t}) (dotted curves), and the
solutions for the resummation of the smallest mass contributions like in
Eq.~(\ref{pipas42t}) (dashed curves), both expansions from the first up to the
fourth order in the asymptotic expansion. For the pure threshold expansion the
order is indicated explicitly.}
\end{figure}
\begin{eqnarray}\label{pidas42t}
\tilde\rho^{das}(E)&=&\frac{\sqrt{2m_0mE}}{8\pi^2M^{3/2}}
  \Bigg\{1+\left(\frac1m+\frac1{m_0}-\frac7M\right)\frac{E}4\nonumber\\&&
  -\left(\frac1{m_0^2}+\frac1{m^2}+\frac{12}{m_0m}-\frac{79}{M^2}\right)
  \frac{E^2}{32}+O(E^3)\Bigg\}
\end{eqnarray}
($M=m+m_0$, the second order asymptotic expansion should suffice to show the
general features in a short and concise form) with the resummed
result
\begin{eqnarray}\label{pipas42t}
\lefteqn{\tilde\rho^{pas}(E)\ =\ \frac{\sqrt{mE(E+2m_0)}}{8\pi^2(E+M)^{3/2}}
  \Bigg\{{_2F_1}\left(0,\frac12;\frac32;1-\frac{m_0^2}{(E+m_0)^2}\right)}
  \nonumber\\&&+\frac{E(E+2m_0)}{8m(E+M)}\
  {_2F_1}\left(\frac12,1;\frac52;1-\frac{m_0^2}{(E+m_0)^2}\right)\\&&
  -\frac{E^2(E+2m_0)^2}{128m^2(E+M)^2}\left(1+\frac{16m(E+M)}{5(E+m_0)^2}
  \right)\
  {_2F_1}\left(1,\frac32;\frac72;1-\frac{m_0^2}{(E+m_0)^2}\right)+\ldots
  \Bigg\}.\nonumber
\end{eqnarray}
Having used $\eps=0$ for the finite spectral density, the first term in curly
braces of Eq.~(\ref{pipas42t}) is obviously equal to $1$ in this limit
because the first parameter of the hypergeometric function vanishes. However,
I keep Eq.~(\ref{pipas42t}) in its given form to show the structure of the
contributions. The result of the comparison is shown in Fig.~\ref{fig07}. The
convergence of different orders of resummed expansions is remarkable, while
the pure exansion diverges already at an early stage.

\addtocounter{figure}{2}%
\begin{figure}[t]\begin{center}
\epsfig{figure=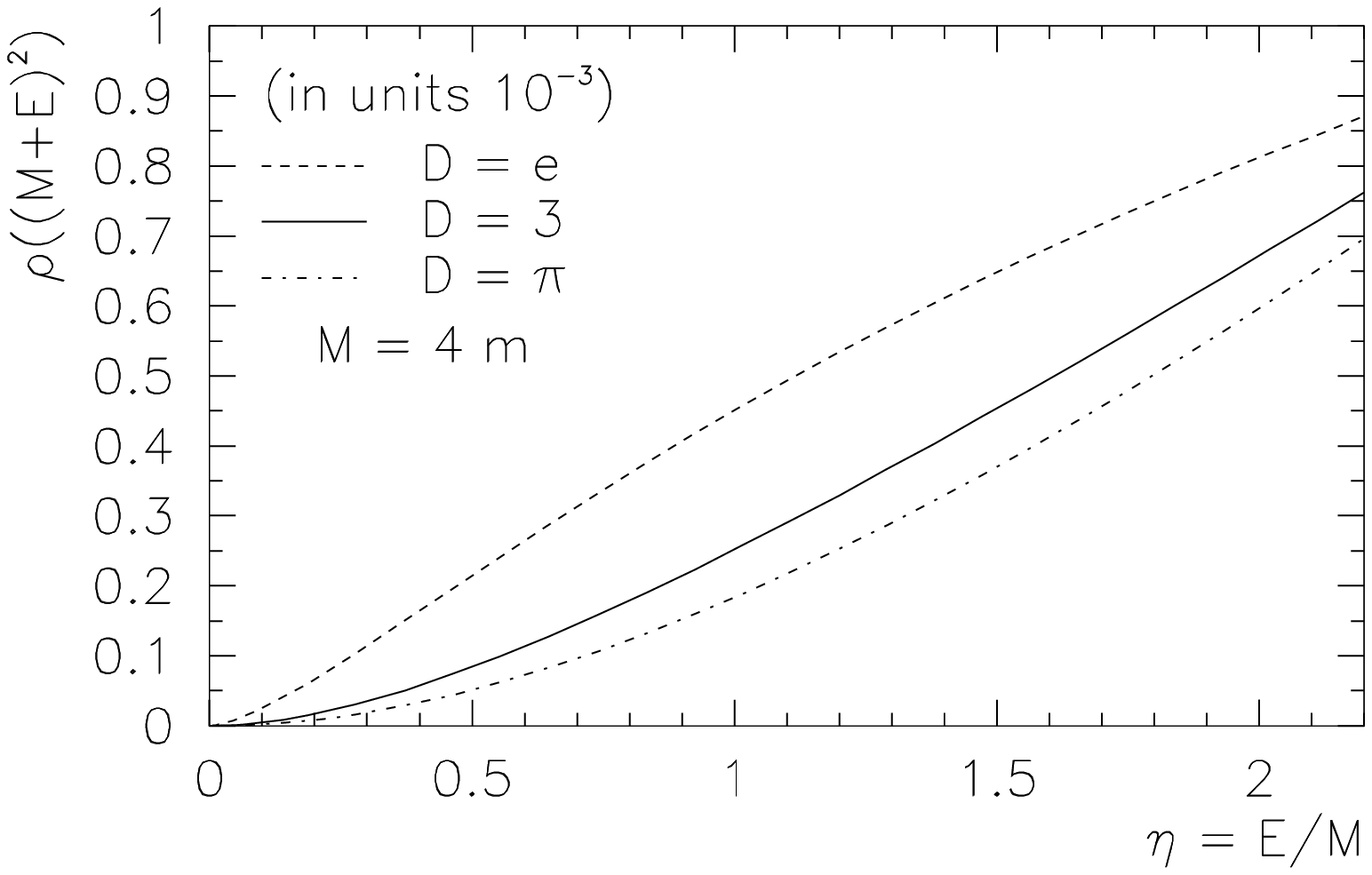,scale=0.8}\end{center}
\caption{\label{fig08}The spectral density for the four-line sunrise-type
diagram with equal masses for $D=e=2.718\ldots$, $D=3$, and $D=\pi=3.14\ldots$
space-time dimensions, to demonstrate the practical convenience of the method.}
\end{figure}

\subsection{A last diagram}
The last figure I want to show is a funny one. In order to demonstrate the
universality of the configuration space method, in Fig.~\ref{fig08} I show
the spectral density for the $n=4$ sunrise-type diagram obtained via the
inverse $K$-transform in Eq.~(\ref{invKtrans}) for the space-time dimensions
$D=e=2.718\ldots$, $D=3$, and $D=\pi=3.14\ldots$.

\newpage

\subsection*{Acknowledgements}
At this point I want to thank the organizers of this conference for the
interesting and challenging opportunity to give a lecture series about the
subject I'm already working on since long. I enjoyed the interesting comments
and discussions during or after my lectures and the stimulating atmosphere of
the conference in general. I want to thank my collaborators J\"urgen K\"orner
and Alexei Pivovarov for the ongoing work on this subject. My work is
supported by the DFG, Germany via the Graduiertenkolleg ``Eichtheorien'' in
Mainz.


\begin{thebibliography}{999}

\bibitem{Grozin:2002nb}A.G.~Grozin,
  Nucl.~Instrum.~Meth.\ {\bf A502} (2003) 815

\bibitem{Mastrolia:2002gt}P.~Mastrolia and E.~Remiddi,
``Analytic evaluation of Feynman graph integrals'',
  [arXiv:hep-ph/0211210]

\bibitem{Mastrolia:2002tv}P.~Mastrolia and E.~Remiddi,
  Nucl.~Phys.\ {\bf B657} (2003) 397

\bibitem{Czyz:2002re}H.~Czy\.z, A.~Grzeli\'nska and R.~Zabawa,
  Phys.~Lett.\ {\bf B538} (2002) 52

\bibitem{Schroder:2002re}Y.~Schr\"oder,
``Automatic reduction of four-loop bubbles'',
  [arXiv:hep-ph/0211288]

\bibitem{Berends:1993ee}F.A.~Berends, M.~Buza, M.~B\"ohm and R.~Scharf,
  Z.~Phys.\ {\bf C63} (1994) 227

\bibitem{Post:1997dk}P.~Post and K.~Schilcher,
  Phys.~Rev.~Lett.\ {\bf 79} (1997) 4088

\bibitem{Post:1996gg}P.~Post and J.B.~Tausk,
  Mod.~Phys.~Lett.\ {\bf A11} (1996) 2115

\bibitem{Rajantie:1996cw}A.K.~Rajantie,
  Nucl.~Phys.\ {\bf B480} (1996) 729; {\bf B513} (1998) 761(E)

\bibitem{Berends:1997vk}F.A.~Berends, A.I.~Davydychev and N.I.~Ussyukina,\\
  Phys.~Lett.\ {\bf B426} (1998) 95

\bibitem{Gasser:1998qt}J.~Gasser and M.E.~Sainio,
  Eur.~Phys.~J.\ {\bf C6} (1999) 297

\bibitem{Mendels:wc}E.~Mendels,
  Nuovo~Cim.\ {\bf A45} (1978) 87

\bibitem{Groote:1998ic}
S.~Groote, J.G.~K\"orner and A.A.~Pivovarov,
  Phys.~Lett.\ {\bf B443} (1998) 269

\bibitem{Groote:1998wy}
S.~Groote, J.G.~K\"orner and A.A.~Pivovarov,
  Nucl.~Phys.\ {\bf B542} (1999) 515

\bibitem{Groote:1999cx}
S.~Groote, J.G.~K\"orner and A.A.~Pivovarov,
  Eur.~Phys.~J.\ {\bf C11} (1999) 279

\bibitem{Groote:2000kz}
S.~Groote and A.A.~Pivovarov,
  Nucl.~Phys.\ {\bf B580} (2000) 459

\bibitem{Mendels:qe}E.~Mendels,
  J.~Math.~Phys.\ {\bf 43} (2002) 3011

\bibitem{Bashir:2001ad}A.~Bashir, R.~Delbourgo and M.L.~Roberts,
  J.~Math.~Phys.\ {\bf 42} (2001) 5553

\bibitem{Delbourgo:2003zi}R.~Delbourgo and M.L.~Roberts,
  J.~Phys.\ {\bf A36} (2003) 1719

\bibitem{Caffo:2002wm}M.~Caffo, H.~Czy\.z and E.~Remiddi,
``Numerical evaluation of master integrals from differential equations'',
  [arXiv:hep-ph/0211178]

\bibitem{Caffo:2002ch}M.~Caffo, H.~Czy\.z and E.~Remiddi,
  Nucl.~Phys.\ {\bf B634} (2002) 309

\bibitem{Ligterink:1999mu}N.E.~Ligterink,
  Phys.~Rev.\ {\bf D61} (2000) 105010;\\
B.~Kastening and H.~Kleinert,
  Phys.~Lett.\ {\bf A269} (2000) 50

\bibitem{Fleischer:1999mp}J.~Fleischer and M.Y.~Kalmykov,
  Phys.~Lett.\ {\bf B470} (1999) 168

\bibitem{Broadhurst:1991fi}D.J.~Broadhurst,
  Z.~Phys.\ {\bf C54} (1992) 599

\bibitem{Groote:1999cn}S.~Groote, J.~G.~K\"orner and A.~A.~Pivovarov,
  Phys.~Rev.\ {\bf D60} (1999) 061701

\bibitem{Laporta:2001dd}S.~Laporta,
  Int.~J.~Mod.~Phys.\ {\bf A15} (2000) 5087

\bibitem{Passarino:2001wv}G.~Passarino,
  Nucl.~Phys.\ {\bf B619} (2001) 257

\bibitem{Suzuki:2000wj}A.T.~Suzuki and A.G.~Schmidt,
  J.~Comput.~Phys.\ {\bf 168} (2001) 207

\bibitem{Witten:1979kh}E.~Witten,
  Nucl.~Phys.\ {\bf B160} (1979) 57

\bibitem{Davydychev:2000na}A.I.~Davydychev and M.Y.~Kalmykov,
  Nucl.~Phys.\ {\bf B605} (2001) 266

\bibitem{Chetyrkin:1996ia}K.G.~Chetyrkin, J.H.~K\"uhn and A.~Kwiatkowski,
  Phys. Rep. {\bf 277} (1996) 189 

\bibitem{Groote:1999zp}S.~Groote, J.G.~K\"orner and A.A.~Pivovarov,
  Phys.~Rev.\ {\bf D61} (2000) 071501;\\
  B.L.~Ioffe,
  Nucl.~Phys.\ {\bf B188} (1981) 317; {\bf B191} (1981) 591(E)

\bibitem{Pivovarov:1991nk}A.A.~Pivovarov and L.R.~Surguladze,
  Nucl.~Phys.\ {\bf B360} (1991) 97,
  Yad.~Fiz.\ {\bf 48} (1988) 1856
  [Sov.~J.~Nucl.~Phys.\ {\bf 48} (1989) 1117]

\bibitem{Ovchinnikov:1991mu}
  A.A.~Ovchinnikov, A.A.~Pivovarov and L.R.~Surguladze,
  Int.~J.~Mod.~Phys.\ {\bf A6} (1991) 2025,
  Sov.~J.~Nucl.~Phys.\ {\bf 48} (1988) 358
  [Yad.~Fiz.\ {\bf 48} (1988) 562]

\bibitem{Groote:2000py}S.~Groote, J.G.~K\"orner and A.A.~Pivovarov,
``Analytical calculation of heavy baryon correlators in NLO of
  perturbative QCD'',
  [arXiv:hep-ph/0009218]

\bibitem{Grozin:1992td}A.G.~Grozin and O.I.~Yakovlev,
  Phys.~Lett.\ {\bf B285} (1992) 254

\bibitem{Furnstahl:1995nd}R.J.~Furnstahl, X.m.~Jin and D.B.~Leinweber,
  Phys.~Lett.\ {\bf B387} (1996) 253

\bibitem{Jin:1997pb}X.m.~Jin and J.~Tang,
  Phys.~Rev.\ {\bf D56} (1997) 515

\bibitem{Kras}N.V.~Krasnikov, A.A.~Pivovarov and A.N.~Tavkhelidze,
  Z.~Phys.\  {\bf C19} (1983) 301;
  JETP Lett.\  {\bf 36} (1982) 333
  [Pisma Zh.~Eksp.~Teor.~Fiz.\ {\bf 36} (1982) 272]

\bibitem{Groote:2001vr}S.~Groote and A.A.~Pivovarov,
  Eur.~Phys.~J.\ {\bf C21} (2001) 133;\\
  JETP Lett.\ {\bf 75} (2002) 221
  [Pisma Zh.~Eksp.~Teor.~Fiz.\ {\bf 75} (2002) 267]

\bibitem{Kataev:1982gr}A.L.~Kataev, N.V.~Krasnikov and A.A.~Pivovarov,\\
  Nucl.~Phys.\ {\bf B198} (1982) 508; {\bf B490} (1997) 505(E)
  Phys.~Lett.\ {\bf B107} (1981) 115

\bibitem{Pivovarov:1999mr}A.A.~Pivovarov,
  Phys.~Atom.~Nucl.\ {\bf 63} (2000) 1646
  [Yad.~Fiz.\ {\bf 63N9} (2000) 1734]

\bibitem{Kazakov:km}D.I.~Kazakov,
  Phys.~Lett.\ {\bf B133} (1983) 406

\bibitem{Bardin:1994sc}D.Y.~Bardin, M.S.~Bilenky, D.Lehner, A.~Olchevski
and T.~Riemann,\\
  Nucl.\ Phys.\ Proc.\ Suppl.\ {\bf 37B} (1994) 148

\bibitem{Wetzorke:2002mx}I.~Wetzorke and F.~Karsch,
``The H dibaryon on the lattice'',
  [arXiv:hep-lat/0208029]

\bibitem{Narison:1994zt}S.~Narison and A.A.~Pivovarov,
  Phys.~Lett.\ {\bf B327} (1994) 341

\bibitem{Weinberg:1978kz}S.~Weinberg,
  Physica {\bf A96} (1979) 327

\bibitem{Larin:1986yt}
  S.A.~Larin, V.A.~Matveev, A.A.~Ovchinnikov and A.A.~Pivovarov,\\
  Yad.~Fiz. {\bf 44} (1986) 1066
  [Sov.~J.~Nucl.~Phys. {\bf 44} (1986) 690];\\
  I.I.~Balitsky, D.I.~D'Yakonov and A.V.~Yung,
  Phys.~Lett.\ {\bf B112} (1982) 71

\bibitem{Sakai:1999qm}T.~Sakai, K.~Shimizu and K.~Yazaki,
  Prog.~Theor.~Phys.~Suppl.\ {\bf 137} (2000) 121

\bibitem{effpot}S.R.~Coleman and E.~Weinberg,
 Phys.~Rev.\ {\bf D7} (1973) 1888;\\
  R.~Jackiw,
  Phys.~Rev.\ {\bf D9} (1974) 1686;\\
  J.M.~Chung and B.K.~Chung,
  J.~Korean Phys.~Soc.\ {\bf 39} (2001) 971;\\
  Phys.~Rev.\ {\bf D59} (1999) 105014

\bibitem{Jackiw:1980kv}
R.~Jackiw and S.~Templeton,
  Phys.~Rev.\ {\bf D23} (1981) 2291

\bibitem{Kajantie:2003ax}
  K.~Kajantie, M.~Laine, K.~Rummukainen and Y.~Schr\"oder,
  JHEP {\bf 0304} (2003) 036

\bibitem{hotQCD}
  T.~Hatsuda,
  Nucl.~Phys.\ {\bf A544} (1992) 27;\\
  T.~Appelquist and R.D.~Pisarski,
  Phys.~Rev.\ {\bf D23} (1981) 2305

\bibitem{Gross:1980br}D.J.~Gross, R.D.~Pisarski and L.G.~Yaffe,
  Rev.~Mod.~Phys.\ {\bf 53} (1981) 43

\bibitem{Nishikawa:2003js}T.~Nishikawa, O.~Morimatsu and Y.~Hidaka,\\
``On the thermal sunset diagram for scalar field theories'',
  arXiv:hep-ph/0302098

\bibitem{Andersen:2000zn}J.O.~Andersen, E.~Braaten and M.~Strickland,
  Phys.~Rev.\ {\bf D62} (2000) 045004

\bibitem{nuclph}
  J.F.~Yang, J.~Zhou and C.~Wu,
``Numerical evaluation of a two loop diagram in the cut off regularization'',
  [arXiv:hep-ph/0301205];
  H.~Van Hees and J.~Knoll,
  Phys.~Rev.\ {\bf D65} (2002) 105005;
  N.P.~Mehta, C.~Felline, J.R.~Shepard and J.~Piekarewicz,
``Low-energy operators in effective theories'',
  [arXiv:nucl-th/0305007]

\bibitem{Platter:2002yr}L.~Platter, H.W.~Hammer and U.G.~Mei\ss ner,
  Nucl.~Phys.\ {\bf A714} (2003) 250


\bibitem{Watson}G.N.~Watson,
  ``Theory of Bessel functions'', Cambridge, 1944

\bibitem{Chetyrkin:pr}K.G.~Chetyrkin, A.L.~Kataev and F.V.~Tkachov,
  Nucl.~Phys.\ {\bf B174} (1980) 345

\bibitem{Terrano:1980af}A.E.~Terrano,
  Phys.~Lett.\ {\bf B93} (1980) 424

\bibitem{Bogoliubov}N.N.~Bogoliubov and D.V.~Shirkov,
  ``Quantum fields'', Benjamin, 1983

\bibitem{Pivovarov:re}A.A.~Pivovarov,
  Phys.~Lett.\ {\bf B236} (1990) 214;
  Phys.~Lett.\ {\bf B263} (1991) 282

\bibitem{Prudnikov}A.P.~Prudnikov, Yu.A.~Brychkov and O.I.~Marichev,\\
  ``Integrals and Series'', Vol.~2, Gordon and Breach, New York, 1990

\bibitem{Gradshteyn}I.S.~Gradshteyn and I.M.~Ryzhik,\\
  ``Tables of integrals, series, and products'', Academic Press, 1994

\bibitem{SonDo}Hoang Son Do, Ph.~D.\ thesis, 2003

\bibitem{Broadhurst:1998rz}D.J.~Broadhurst,
  Eur.~Phys.~J.\ {\bf C8} (1999) 311

\bibitem{Pivovarov:ij}A.A.~Pivovarov, N.N.~Tavkhelidze and V.F.~Tokarev,
  Phys.~Lett.\ {\bf B132} (1983) 402

\bibitem{Chetyrkin:yr}K.G.~Chetyrkin and A.A.~Pivovarov,
  Nuovo Cim.\ {\bf A100} (1988) 899

\bibitem{Pivovarov:jm}A.A.~Pivovarov and V.F.~Tokarev,
  Yad.~Fiz.\ {\bf 41} (1985) 524

\bibitem{Meijer}C.S.~Meijer, Proc.~Amsterdam~Akad.~Wet.\ (1940) 599; 702

\bibitem{Erdelyi}A.~Erdelyi (Ed.), ``Tables of integral transformations'',\\
  Volume~2, Bateman manuscript project, 1954

\bibitem{AbramowitzStegun}M.~Abramowitz, I.A.~Stegun (eds.), ``Handbook of
  Mathematical Functions'',\\ Dover Publ.~Inc., New York, 9th Printing, 1970

\end{thebibliography}
\end{document}